\documentclass[aip,jcp,reprint,superscriptaddress]{revtex4-2}

\usepackage{graphicx}

\usepackage{amsmath,amssymb}

\usepackage{newtxtext,newtxmath}

\usepackage{bm}

\usepackage{booktabs}

\usepackage{xcolor}

\usepackage[
 colorlinks=true,
 linkcolor=blue,
 citecolor=blue,
 urlcolor=blue
]{hyperref}

\providecommand{\ave}[1]{\left\langle#1\right\rangle}
\providecommand{\kww}{_\mathrm{KWW}}

\begin{document}

\title{
Classification of interfacial water governed by water-polymer
interactions in hydrated polymers: A molecular dynamics simulation study
of ethylene-based and acrylate polymers
}

\author{Atsuki Hashimoto}
\affiliation{Division of Chemical Engineering, Department of Materials Engineering Science, Graduate School of Engineering Science, The University of Osaka, Toyonaka, Osaka 560-8531, Japan}

\author{Kokoro Shikata}
\affiliation{Division of Chemical Engineering, Department of Materials Engineering Science, Graduate School of Engineering Science, The University of Osaka, Toyonaka, Osaka 560-8531, Japan}

\author{Kang Kim}
\email{kk@cheng.es.osaka-u.ac.jp}
\affiliation{Division of Chemical Engineering, Department of Materials Engineering Science, Graduate School of Engineering Science, The University of Osaka, Toyonaka, Osaka 560-8531, Japan}

\author{Nobuyuki Matubayasi}
\email{nobuyuki@cheng.es.osaka-u.ac.jp}
\affiliation{Division of Chemical Engineering, Department of Materials Engineering Science, Graduate School of Engineering Science, The University of Osaka, Toyonaka, Osaka 560-8531, Japan}

\date{\today}

\begin{abstract}
We perform molecular dynamics simulations to investigate hydration
 structures and dynamics in seven water-containing polymers: poly(vinyl
 alcohol) (PVA), poly(2-hydroxyethyl acrylate) (PHEA),
 poly(2-hydroxyethyl methacrylate) (PHEMA), poly(butyl acrylate) (PBA),
 poly(2-methoxyethyl methacrylate) (PMEMA), poly(ethylene glycol) (PEG),
 and poly(2-methoxyethyl acrylate) (PMEA). 
The analysis integrates four perspectives: the water-content dependence
 of the glass transition temperature $T_\mathrm{g}$, polymer chain
 fluctuations characterized by dihedral angle distributions,
 hydrogen-bond lifetimes $\tau_{\mathrm{HB}}$ between water and polymer
 functional groups, and the localization and exchange dynamics of
 confined water quantified by the distinct part of van Hove correlation
 function.
Hydroxyl-containing polymers (PVA, PHEA, and PHEMA) exhibit relatively
 high dry-state $T_\mathrm{g}$ values and its pronounced depression upon
 hydration. 
Chain fluctuations are limited, and $\tau_{\mathrm{HB}}$ follows
 Arrhenius behavior, forming localized hydration shells. 
In contrast, PMEMA and PBA show low equilibrium water contents and
 hydrophobic character; although their dry-state $T_\mathrm{g}$ values
 are moderately lower and less sensitive to water content, chain
 fluctuations remain small, and $\tau_{\mathrm{HB}}$ also obeys Arrhenius
 behavior, with hydrophobic aggregation promoting water localization.
PEG and PMEA display low dry-state $T_\mathrm{g}$ values and weak
 water-content dependence. 
Greater rotational freedom around ether or methoxy oxygen atoms leads to
 larger chain fluctuations and loosely bound water. 
Below $T_\mathrm{g}$, $\tau_{\mathrm{HB}}$ between water and ether or
 methoxy oxygen atoms exhibits super-Arrhenius behavior. 
These results clarify three hydration types: highly hydrated (PVA, PHEA,
 and PHEMA),
 hydrophobic (PMEMA and PBA), and flexibly hydrated (PEG and PMEA), and provide a molecular-level
 framework for interpreting 
interfacial water governed by water-polymer interactions.
\end{abstract}

\maketitle

\section{INTRODUCTION}
\addcontentsline{toc}{section}{INTRODUCTION}

\begin{figure*}[t]
\centering
\includegraphics[width=0.8\linewidth]{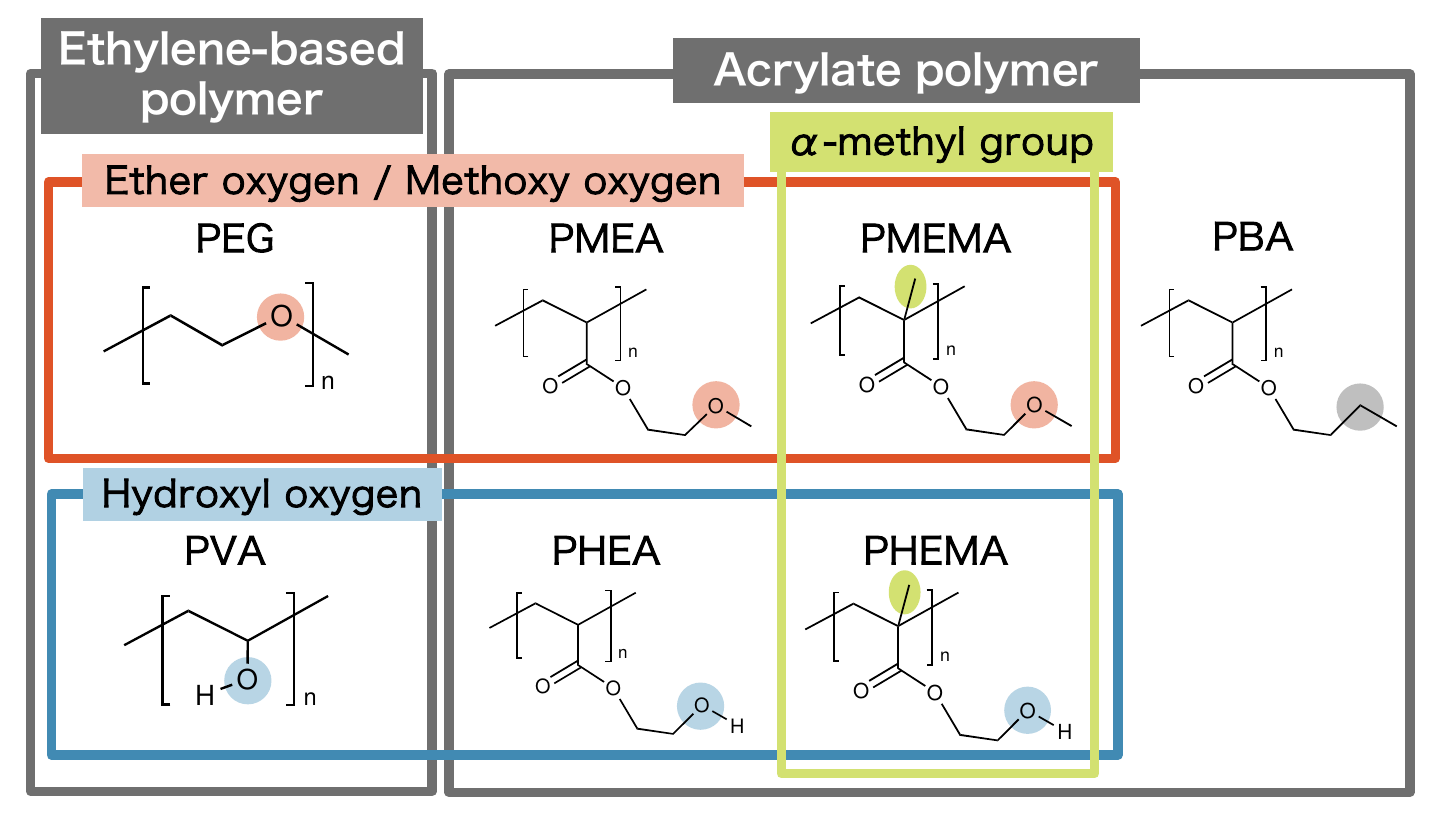}
\caption{Structure and classification of polymers studied in this paper.}
\label{fig:Structure}
\end{figure*}

Water molecules at polymer interfaces in aqueous environments play a
central role in determining polymer functionality. 
Under hydrated
conditions, polymers exhibit behaviors such as swelling, which enables 
functional applications including hydrogels for drug delivery and
contact lenses, temperature-responsive materials, and membranes for
selective separation.~\cite{bayliss2023Hydrophilic}
Interfacial water 
is located near polymer surfaces
through intermolecular
forces such as van der Waals interactions, electrostatic interactions,
and hydrogen bond (H-bond) with functional groups or surface molecules. 
Owing to its distinct microscopic structure and physical chemistry
properties, interfacial water displays behavior that differs
significantly from that of bulk water. 
Consequently, investigations of interfacial water are important for
developing a detailed understanding of water behavior near polymer 
surfaces and for enabling appropriate control of these interfacial
properties.~\cite{wan2025Interfacial}

Interfacial water weakly associated with polymer surfaces is
thought to contribute to the functional performance of biocompatible
polymers.~\cite{tsuruta2010Role, tanaka2026Interfacial}
The scenario can be summarized as follows.
Water molecules that are strongly associated with hydrophilic polymer
surfaces can disrupt the hydration layer that stabilizes protein
structure when proteins come into contact with the material. 
This disruption promotes protein denaturation and aggregation. 
Cells subsequently adhere to these protein aggregates, eventually
leading to thrombus formation. 
In contrast, polymers with high biocompatibility retain water molecules
that are only weakly associated with the interface. 
These interfacial water molecules form a barrier that helps maintain the
hydration layer of proteins at the polymer surface, thereby supporting
biocompatible behavior.

A representative biocompatible polymer is polyethylene glycol (PEG), a
water-soluble polymer synthesized from ethylene oxide.~\cite{harris1992PolyEthylene} 
In addition to its high solubility in organic solvents, PEG exhibits low
toxicity, chemical stability, and lubricity. 
Owing to its ability to suppress protein adsorption and cell adhesion,
as well as its stable dispersion in aqueous media, PEG is widely
employed in a range of applications. 
PEG-immobilized surfaces form highly hydrated layers in aqueous
environments due to their strong affinity for water. 
The resulting hydration layer introduces an energetic barrier to protein
adsorption, thereby inhibiting nonspecific protein binding.~\cite{jeon1991Protein}

Poly(2-methoxyethyl acrylate) (PMEA) is a biocompatible polymer developed by Tanaka
\textit{et al.}~\cite{tanaka2000Cold, tanaka2000Blood, tanaka2004Effect}
PMEA exhibits properties including water insolubility, suppression of
protein adsorption and denaturation, inhibition of platelet adhesion and
activation, and low toxicity.
Experimental research focusing on the hydration structure within PMEA
have been conducted 
to clarify the mechanisms underlying its biocompatibility.
In particular, differential scanning calorimetry (DSC) studies have
reported that PMEA contains characteristic interfacial
water.~\cite{tanaka2015Design, sato2015Relationship}
Interfacial water is commonly classified into three categories:
free water, which freezes and melts at 0 $^\circ$C; bound water, which
remains nonfreezing even at -100 $^\circ$C; and intermediate water,
which undergoes crystallization and melting at temperatures below the
freezing point.
Furthermore, evaluations of platelet adhesion indicate that the amount
of intermediate water is a key factor governing the manifestation of
antithrombotic effects.
Note that PEG, the biocompatible polymer described
above, has also been reported to contain intermediate water.~\cite{hatakeyama2007Cold}

The freezing and melting behavior of water observed by DSC provides
important insights into the strength of interactions between polymers and
water.
Furthermore, low-temperature crystallization of water in the hydrated
polymer was observed immediately after the glass transition during DSC
heating.~\cite{sato2015Relationship}
This behavior suggests that the emergence of intermediate water is
closely related to the glass transition temperature, $T_\mathrm{g}$.
When considering polymer properties, the glass transition temperature
$T_\mathrm{g}$ is a critical parameter, and the expression of polymer functions
depends strongly on the temperature difference between the ambient
temperature and $T_\mathrm{g}$.
PEG and PMEA, both reported to contain intermediate water, exhibit glass
transition temperatures $T_\mathrm{g}$ below $0\,^\circ\mathrm{C}$ in the dry state.
However, because the glass transition temperature $T_\mathrm{g}$ in the dry state
does not describe water-polymer interactions, understanding molecular
behavior in the vicinity of $T_\mathrm{g}$ under hydrated conditions is essential
for elucidating the properties of polymers.
Note that, experimentally, the dependence of $T_\mathrm{g}$
on water content in water-soluble polymers such as PEG and 
poly(vinyl alcohol) (PVA), a polymer with a structure similar to that of
PEG and a representative gel-forming
material, 
is
important for understanding the degree of plasticization induced by
water.~\cite{rault1995Glass, hodge1996Free, kyritsis1997Dielectric,
yoon1997Endgroup, huang2001Interaction, trotzig2007Structure}

Molecular dynamics (MD) simulations provide a powerful approach for
directly probing the local property of interfacial water in
the vicinity of polymers.
PEG, a hydrophilic polymer with a simple polyether backbone, has been
extensively studied using MD simulations, with investigations addressing
its polymer chain structure, 
hydration water content, 
H-bonds with water molecules, and glass transition behavior.~\cite{smith1993Force, smith2000Molecular,
tasaki1996PolyoxyethyleneWater, tasaki1999Polyoxyethylenecation,
borodin2001Molecular, borodin2002Concentration, dormidontova2002Role,
juneja2010Merging, wu2011Simulated, fuchs2012GROMOS, wu2017Glass,
sponseller2021Solutions, klajmon2023Glass, hoffmann2023Behavior,
ho2024Thermophysical, gudla2024How}
Numerous MD simulations have also been performed on
PVA.~\cite{tamai1996Molecular, tamai1996Molecularb,
tamai1996Moleculara, 
muller-plathe1997Solvation, muller-plathe1998Different,
muller-plathe1998Microscopic, 
muller-plathe1998Diffusion, tamai1998Dynamic, tamai1999Effects,
tamai1999Structure, karlsson2002Physical, 
karlsson2004Molecular, chiessi2005Supercooled, chiessi2007Water, 
zhang2009Microstructure, bermejo2009Influence, wu2010Cooperative, tesei2012Poly, sun2022Structure}
Wu reported the glass transition temperature $T_\mathrm{g}$ of PEG in
the dry state for bulk, film, and isolated chains using MD
simulations.~\cite{wu2017Glass}
Sponseller \textit{et al.} also investigated $T_\mathrm{g}$ of PEG.~\cite{sponseller2021Solutions}
Furthermore, Wu examined $T_\mathrm{g}$ in aqueous PVA systems.~\cite{wu2010Cooperative}

In addition, a number of MD studies have investigated the structural and
dynamical characteristics of interface water confined within
PMEA.~\cite{nagumo2013Computational, nagumo2019Molecular,
nagumo2021Interactions, yasoshima2017DiffusionControlled,
kishinaka2019Molecular, kuo2019Analyses, kuo2020Elucidating,
kuo2020Molecular, kuo2021Effects, 
yasoshima2021Molecular, yasoshima2022Adsorption, ikemoto2022Infrared,
yadav2022Adsorption, 
nagumo2023Influence, shikata2023Revealing, yadav2025Hydration,
yadav2025HydrophobicityGoverned, gemmei-ide2025Weak}
Note that 
the glass transition behavior of PMEA has also been investigated by MD
simulations.~\cite{yasoshima2017DiffusionControlled, gemmei-ide2025Weak}
We recently performed MD simulations of PMEA and structurally
related (meth)acrylate polymers, including poly(2-hydroxyethyl
methacrylate) (PHEMA) and poly(1-methoxymethyl acrylate) (PMC1A), to
elucidate the H-bond dynamics associated with each acceptor
oxygen site for each polymer functional group.~\cite{shikata2023Revealing}
Our results indicate that, upon breakage of H-bonds between water
molecules and the methoxy oxygen of PMEA, the water molecules remain in
close proximity and do not diffuse on the picosecond timescale. 
In contrast, water molecules H-bonded to the hydroxy oxygen of PHEMA and
the methoxy oxygen of PMC1A diffuse immediately following hydrogen-bond
breakages.

In this study, we focus on the observation that not only PMEA but also
PEG is a hydrophilic polymer exhibiting antithrombotic properties and
intermediate water, and accordingly performed MD simulations. 
In addition, PVA, poly(2-hydroxyethyl acrylate)
(PHEA), PHEMA, poly(2-methoxyethyl
methacrylate) (PMEMA), and poly(butyl acrylate) (PBA) are simulated as
negative controls. 
This MD analysis enables a comparative, comprehensive examination of
ethylene-based polymers (PEG and PVA) and acrylate polymers (PMEA, PHEA,
PMEMA, PHEMA, and PBA), allowing assessment of how variations in
functional oxygen groups and the presence or absence of methyl side
chains influence the behavior of water molecules in the vicinity of the
polymers.

Figure~\ref{fig:Structure} presents the chemical structures of the seven
polymers examined. 
PEG contains an ether oxygen, whereas PMEA contains a carbonyl
oxygen, an ether-like oxygen, and a methoxy oxygen. 
Replacement of the ether group with a hydroxyl group yields PVA and
poly(2-hydroxyethyl acrylate) (PHEA), respectively. 
Introduction of an $\alpha$-methyl side chain on the polymer backbone
gives poly(2-methoxyethyl methacrylate) (PMEMA) and PHEMA, respectively,
while replacement of the methoxy 
group in PMEA with an ethyl group results in poly(butyl acrylate) (PBA).
The water solubility, equilibrium water content, glass transition
temperature $T_\mathrm{g}$ in the dry state, blood compatibility, and presence or
absence of intermediate water for these polymers are summarized in
Table~S1 of the Supplementary Material.

The analyses conducted in this study are summarized as follows. 
First,
the glass transition temperature $T_\mathrm{g}$ under both dry and hydrated conditions was
determined from the temperature dependence of the mass density.
To assess the mobility of hydrated polymer chains, fluctuations in
dihedral angle conformations were then examined. 
In addition, the H-bonding state of water molecules was characterized,
and the associated time scales were quantified as H-bond lifetimes, with
their temperature dependence analyzed. 
In combination, these analyses provide insight into water-polymer
interactions. 
Finally, to clarify the mobility of water molecules confined within the
polymers, the distinct part of van Hove correlation function, which describes molecular
exchange processes among water molecules, was evaluated.

\begin{figure}[t]
\centering
\includegraphics[width=0.9\linewidth]{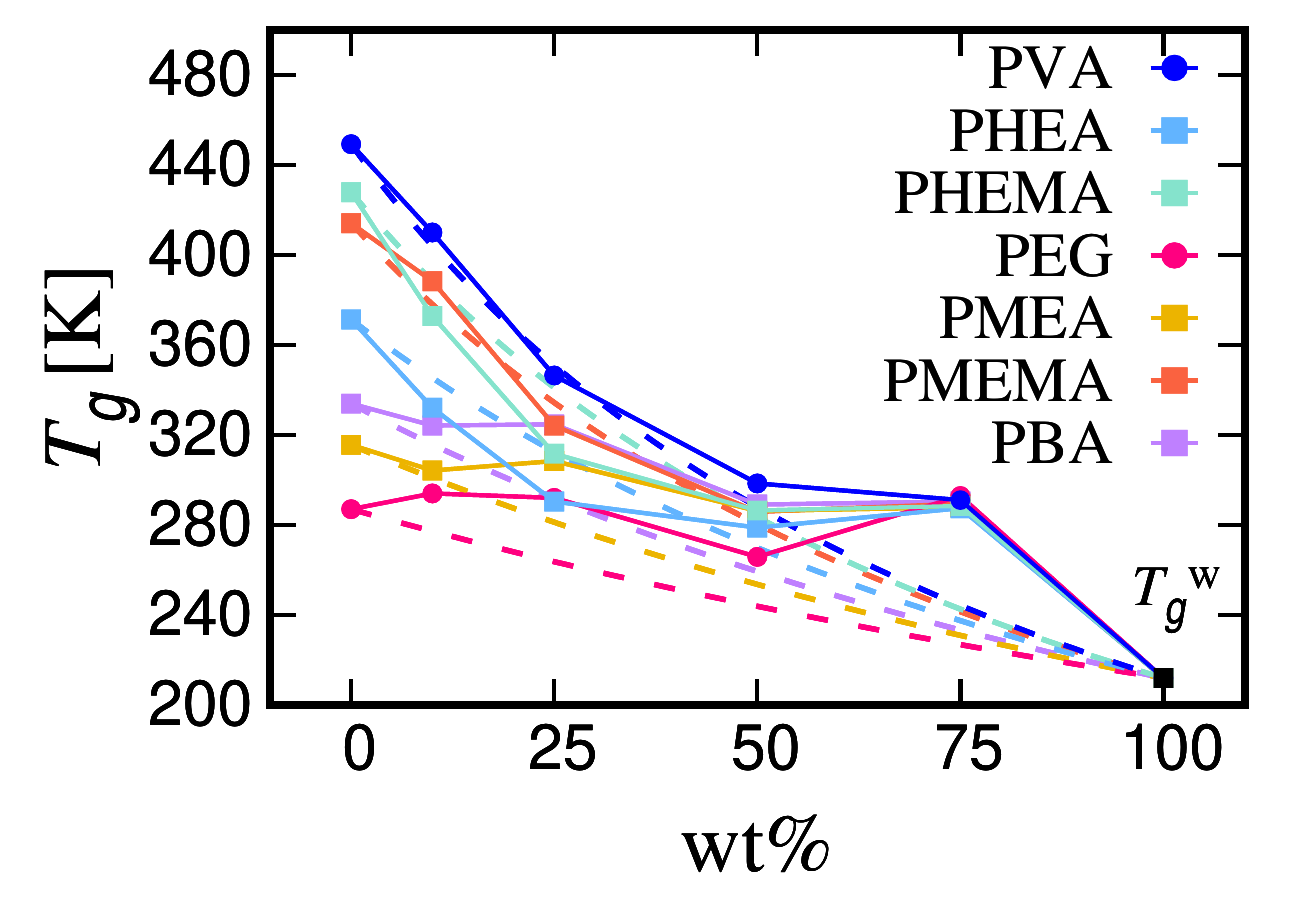}
\caption{Water content dependence of the glass transition temperature
 $T_\mathrm{g}$. 
$T_\mathrm{g}^\mathrm{w}\approx 212$ K denotes the glass transition temperature of the
 TIP4P/2005 water model, as previously
 reported.~\cite{kreck2014Characterization}
The dashed line represents the Fox equation [Eq.~\eqref{eq:Fox}], which estimates $T_\mathrm{g}$ for
 the water-polymer mixture system based on the glass transition
 temperatures of the dry
 polymer and pure water.}
\label{fig:tg}
\end{figure}

\section{SIMULATION DETAILS}

Each polymer was modeled using J-OCTA~\cite{JOCTA} with the OPLS-AA
force field.~\cite{jorgensen1996Development} 
The
polymer chains were constructed as atactic with a 1:1 stereoregularity
and a degree of polymerization of 50. 
Additionally, quantum chemistry calculations were performed to determine atomic charges 
at a degree of oligomerization of 11 with the same form of tacticity, 
following the protocol
described in Ref.~\onlinecite{kojima2021Water}. 
H atoms were added to both polymer terminal, and
the number of polymer molecules in the simulation box was set to
20. 
Water molecules were represented by the TIP4P/2005
model~\cite{abascal2005General} and added to the systems 
under periodic boundary conditions to achieve the mass water contents
of 0, 10, 25, 50, and 75 wt\%. 
MD simulations were
carried out using GROMACS 2024.4,~\cite{abraham2015GROMACS} and the initial configurations were
generated with PACKMOL.~\cite{martinez2009PACKMOL}
It has been noted that caution is required when applying the
OPLS-AA force field to PEG.~\cite{hoffmann2023Behavior}
However, in this study, we employ the OPLS-AA force field to analyze all
seven polymers in a consistent manner.

After energy minimization, system equilibration was performed in a
manner similar to 
that described in 
Ref.~\onlinecite{kojima2021Water}, with details provided in
Table~S2 of the Supplementary Material.
$NVT$ simulations employed the velocity-rescaling method for temperature
control.~\cite{bussi2007Canonical} 
In $NPT$ simulations, temperature control was also achieved using velocity
rescaling, while pressure was regulated using the stochastic
cell-rescaling method.~\cite{bernetti2020Pressure}
In all MD simulations, the time step was set to 1 fs.
MD Simulation snapshots are illustrated in Fig.~S1 of the
Supplementary Material.

\begin{figure}[t]
\centering
\includegraphics[width=\linewidth]{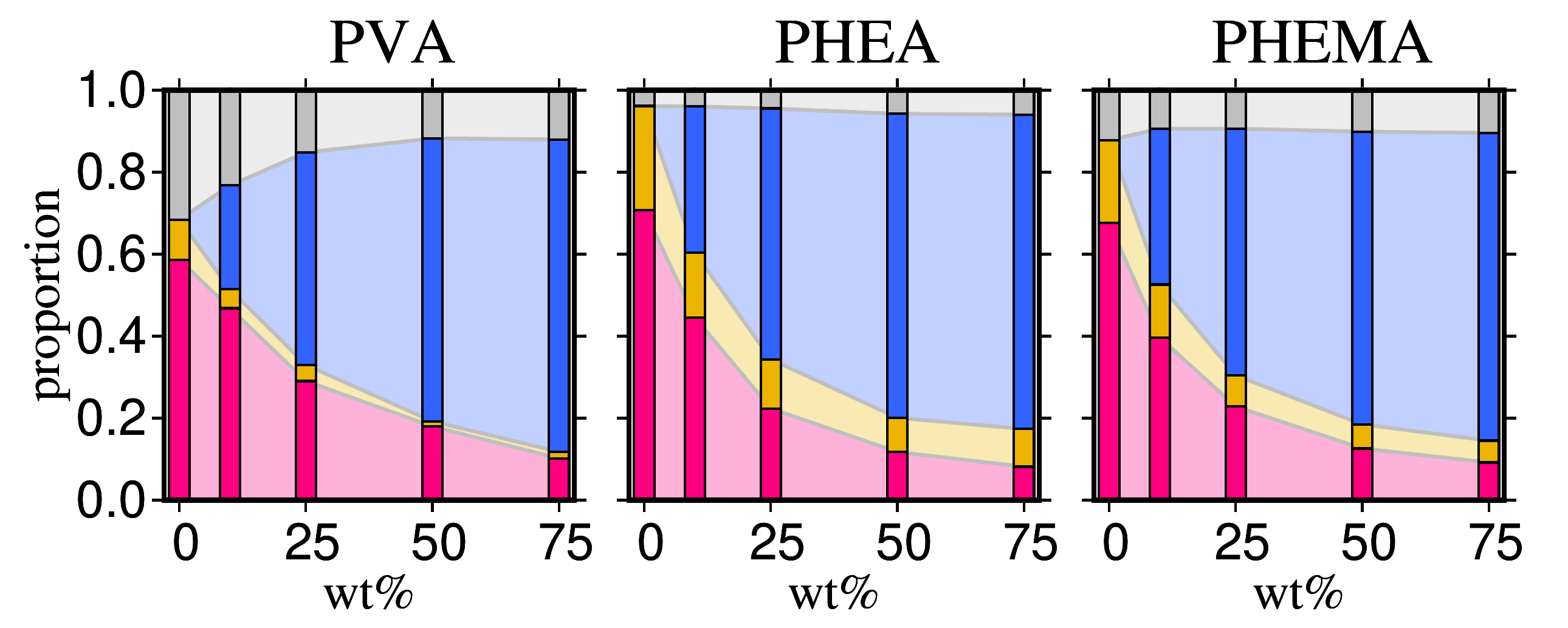}
\caption{Proportions of H-bonding partners for hydroxyl oxygen
 atoms in PVA, PHEA, and PHEMA at 300 K. 
Magenta denotes intermolecular H-bonds to oxygen acceptors on other
 polymer chains; yellow denotes intramolecular H-bonds to oxygen atoms
 on the same polymer chain; blue denotes H-bonds to oxygen atoms of
 water molecules; gray denotes hydroxyl oxygen atoms that do not participate in H-bonding.}
\label{fig:numhb-poh}
\end{figure}

\section{RESULTS AND DISCUSSION}

\subsection{Glass transition temperature}
\label{section:tg}

To examine the temperature dependence of the mass density $\rho$, MD
simulations were conducted during a cooling protocol. 
After equilibration at 600 K (for 0 wt\% and 10 wt\% systems of PVA, PHEA, PMEMA, and
PHEMA) or 500 K (for the other systems), the temperature was reduced
stepwise by 20 K, and at each temperature, 
$NPT$ simulations of 4.05 ns followed by
$NVT$ simulations of 10 ns were performed.
Figure~S2 of the Supplementary Material presents the
temperature dependence of $\rho$ for each polymer at different water
contents.
It also presents the temperature dependence of the mass density of pure
water described by the TIP4P/2005 model using a system of 1000 molecules.
For the pure water system, 
the temperature dependence of the mass density exhibits nonmonotonic
behavior, with a maximum at approximately 280 K followed by an
increase at lower temperatures, as shown in Fig.~S2 of
the Supplementary Material.
This behavior reflects the presence of two liquid states of water,
namely, a
high-density liquid and a low-density
liquid.~\cite{singh2016Twostate, biddle2017Twostructure, saito2018Crucial}
Kreck \textit{et al.} characterized the glass transition temperature
$T_\mathrm{g}$ of various water models by calculating heat capacity using MD
simulations, reporting $T_\mathrm{g} \approx 212$ K for the 
TIP4P/2005 model.~\cite{kreck2014Characterization}

At a water content of 0 wt\% in the dry state, polymers containing
hydroxyl groups exhibit higher densities than those containing ether
groups, consistent with experimental tendency.
Furthermore, the density decreases with increasing water content for each polymer.
As the temperature decreases, a change in slope is observed at a
characteristic temperature, which defines the glass transition
temperature $T_\mathrm{g}$.
A common approach for estimating the glass transition temperature $T_\mathrm{g}$ is to
approximate the density-temperature dependence with two linear fits and
determine $T_\mathrm{g}$ from their intersection.
Figure~S2 of the Supplementary Material 
shows these linear fits for each polymer at water contents below 25 wt\%.
However, at water contents above 50 wt\%, 
the non-monotonic behavior in density-temperature curve was
observed, reflecting the anomalous temperature dependent
of pure water, as noted above.
In this regime, 
linear fitting becomes difficult; 
therefore, $T_\mathrm{g}$ is defined as the temperature at which the
second derivative of $\rho$ with respect to temperature attains a
minimum.
For this purpose, the Savitzky--Golay filter was applied to perform
second-order differentiation of the density-temperature data.~\cite{savitzky1964Smoothing}
The resulting $T_\mathrm{g}$ determined from the minimum of the second derivative of $\rho$
with respect to $T$ is shown in Fig.~S2 of the
Supplementary Material.

\begin{figure}[t]
\centering
\includegraphics[width=\linewidth]{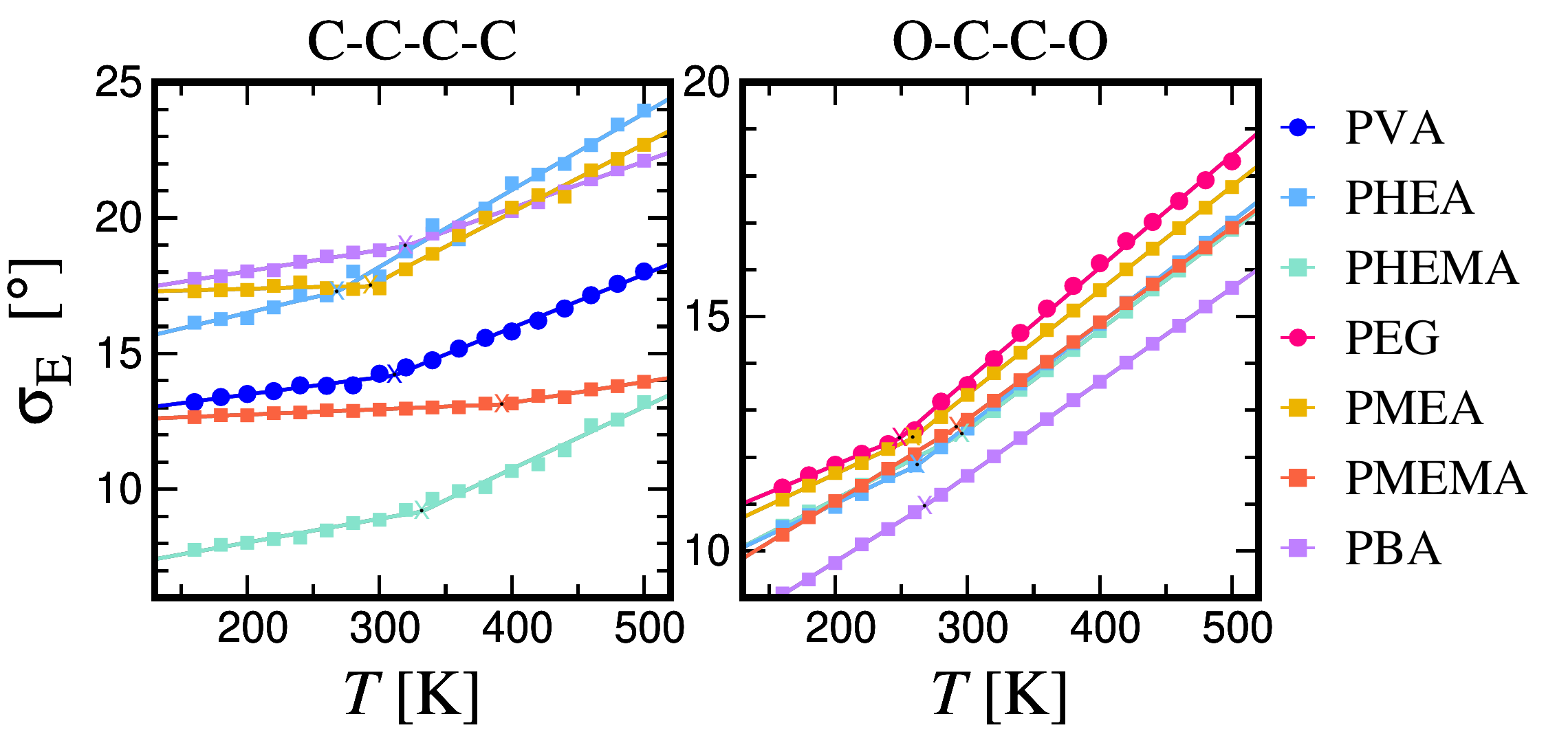}
\caption{Temperature dependence of $\sigma_\mathrm{E}$, which quantifies
 the spread of the Gaussian mixture model distribution, at a water
 content of 25 wt\%. 
The left panel shows the main-chain carbon dihedral angle (C-C-C-C), and
 the right panel shows the carbon skeleton dihedral angle associated
 with the oxygen substituent (O-C-C-O). 
For PBA, the latter corresponds to the O-C-C-C dihedral angle that
 includes the butyl group.
Straight lines denote the linear fits, and the symbol of $\times$
 indicates the intersection that determine the glass transition
 temperature $T_\mathrm{g}^\mathrm{C-C-C-C}$ and $T_\mathrm{g}^\mathrm{O-C-C-O}$ for the 
 C-C-C-C and O-C-C-O dihedral angles, respectively.}
\label{fig:sigma}
\end{figure}

\begin{figure*}[t]
\centering
\includegraphics[width=0.8\linewidth]{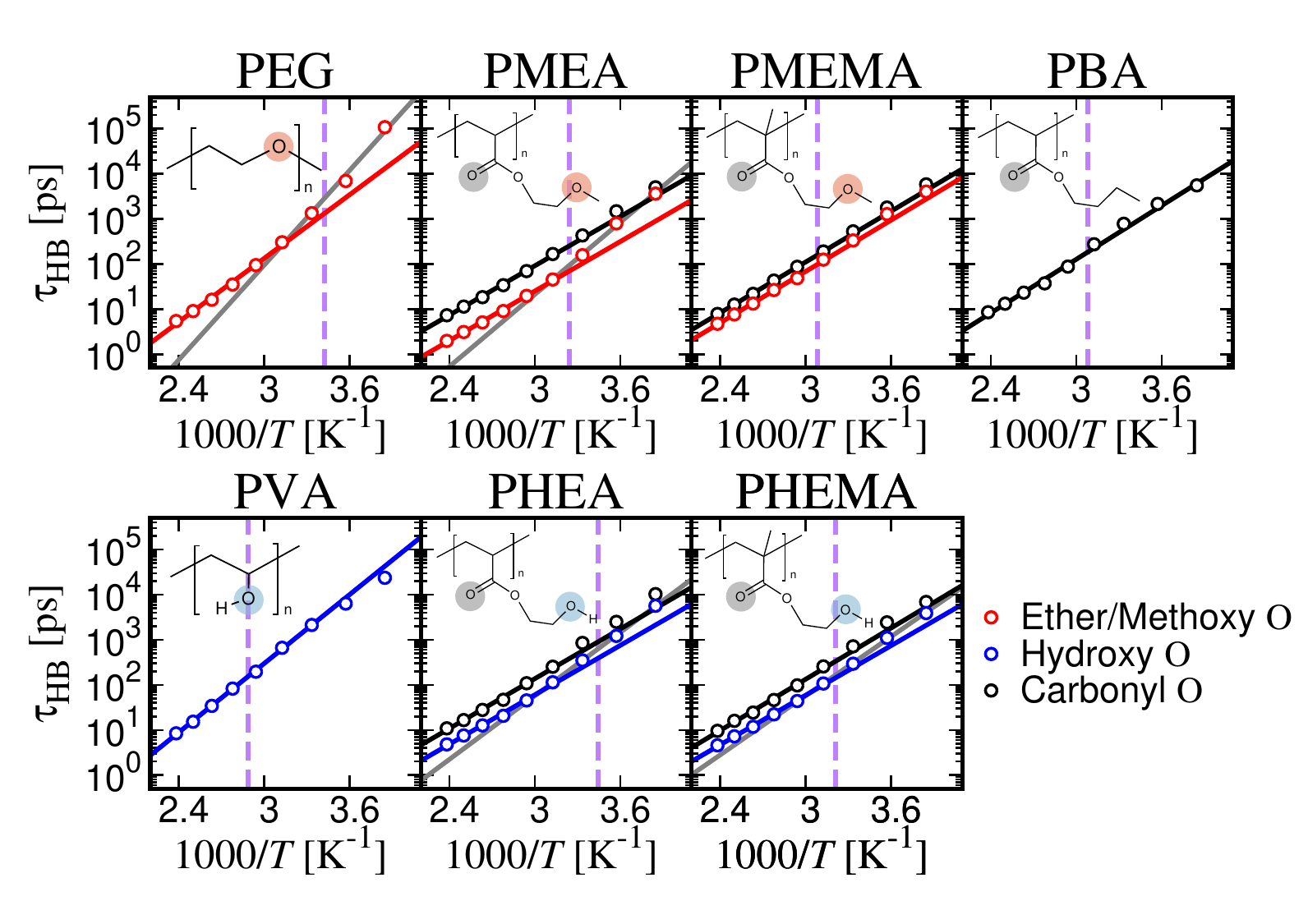}
\caption{Arrhenius plots of $\tau_{\mathrm{HB}}$ between water molecules
 and acceptor oxygens of polymer functional groups at a water content of
 25 wt\% (260-420 K in 20 K intervals). 
Red symbols denote H-bonds with ether or methoxy oxygens, red symbols
 correspond to hydroxyl oxygens, and black symbols represent carbonyl
 oxygens in acrylate groups as acceptors. 
The blue and red lines indicate linear fits using the Arrhenius equation in the high-temperature range
 (340-420 K). 
The gray lines correspond to fits in the low-temperature range (260-320 K).
The purple dashed line represents the estimated glass transition temperature $T_\mathrm{g}$.}
\label{fig:tauhb}
\end{figure*}

Figure~\ref{fig:tg}
shows the glass transition temperature $T_\mathrm{g}$ as a function of the water
content for the seven polymer systems.
For reference, an empirical relation based on the Fox equation for
estimating the glass transition temperature of a binary mixture is also
shown. 
The Fox equation assumes equal free-volume fractions 
for components 1 and 2 and is expressed 
as 
\begin{equation}
\frac{1}{T_\mathrm{g}} = \frac{w_1}{T_{\mathrm{g},1}} + \frac{w_2}{T_{\mathrm{g},2}},
\label{eq:Fox}
\end{equation}
where $w_1$ and $w_2$ are the weight fractions of component 1 and 2,
respectively and $T_{\mathrm{g},1}$ and $T_{\mathrm{g},2}$ denote the
corresponding glass transition
temperatures.~\cite{fox1956Influence}

In the dry state (0 wt\%), MD simulations reproduced consistent trends in the glass
transition temperatures.
However, owing to the high cooling rates employed in the simulations,
the calculated $T_\mathrm{g}$ values are approximately 40-120 K higher than the
experimental values. 
A study by Soldera and Metatla proposed a method for
correcting $T_\mathrm{g}$ values obtained from MD simulations using the
Williams--Landel--Ferry (WLF) equation, 
\begin{equation}
 \Delta T_\mathrm{g} = 
\frac{-B \log_{10} \frac{q_\mathrm{sim}}{q_\mathrm{exp}}}
{A+ \log_{10}\frac{q_\mathrm{sim}}{q_\mathrm{exp}}},
\label{eq:WLF}
\end{equation}
where $\Delta T_\mathrm{g}$ represents the difference in $T_\mathrm{g}$ between cooling rates
$q_\mathrm{sim}$ in MD simulations and $q_\mathrm{exp}$
in experiments.~\cite{soldera2006Glassa}
Furthermore, the constants $A = 16.7$ and $B = 48$ K are fitting
parameters determined as
average values for some polymer systems.
Figure~S3 of the Supplementary Material shows the
relationship between the corrected $T_\mathrm{g}$ values
from MD simulations and the experimental values.
This indicates that the corrected $T_\mathrm{g}$ is consistent with
experimentally reported values, within a deviation of approximately 40 K.
Note that PMEMA shows the largest overestimation from the experimental
values by the MD simulations.
Nevertheless, of the seven polymers examined, PVA, PHEA, and PHEMA exhibit
relatively higher $T_\mathrm{g}$ values in the dry state, consistent with
experimental observations.
This behavior is attributed to the formation of intermolecular H-bonds
mediated by hydroxyl groups and 
to increased chain rigidity arising from $\alpha$-methyl side chains
along the polymer backbone.

However, because experimental $T_\mathrm{g}$ values under water-saturated
conditions are limited, the following discussion is based on uncorrected
$T_\mathrm{g}$ values without applying the WLF equation.
Figure~\ref{fig:tg} demonstrates that 
with addition of water, $T_\mathrm{g}$ reduces from that at the dry state, 
with this trend being
particularly pronounced for polymers with high dry-state $T_\mathrm{g}$. 
This behavior arises from water-induced plasticization of the polymer matrix.~\cite{hancock1994Relationship}
Indeed, a pronounced decrease in the glass transition temperature with
increasing water content has been experimentally observed for
PVA,~\cite{rault1995Glass, hodge1996Free, kyritsis1997Dielectric} PHEA,~\cite{kyritsis1994Dielectric}
and PHEMA.~\cite{verhoeven1989Physicochemical, bouwstra1995Thermal, meakin2003Thermal}
Furthermore, although the Fox equation captures the overall trend, it
underestimates the magnitude of the plasticizing effect, particularly for PHEA and PHEMA.
This suggests that hydroxyl groups in PHEA and PHEMA contribute to
plasticization not only through H-bond formation between polymer chains
but also via H-bonding between the polymer and water molecules. 

To elucidate the plasticization effect of water molecules, particularly
for PVA, PMEA, and PHEMA, we
analyzed in detail the water content dependence of the probability distribution of
H-bonding partners associated with hydroxyl oxygen atoms.
Here, H-bond between donor and acceptor was defined when the oxygen-oxygen distance 
$r_\mathrm{OO}$ was 0.35 nm or less and the
oxygen$\cdots$oxygen-hydrogen angle was 30$^\circ$ or less.~\cite{luzar1996Effect,
luzar1996Hydrogenbond, luzar2000Resolving}
The results obtained at a temperature of 300 K are shown in Fig.~\ref{fig:numhb-poh}.
In the dry state, approximately 90\% of the hydroxyl oxygen atoms in
PHEA and PHEMA participate in H-bonding, and this fraction remains
essentially unchanged upon hydration.
As the water content increased, the fractions of intermolecular and
intramolecular H-bonds decreased, while the proportion of H-bonds
formed with water molecules increased.
Accordingly, in PHEA and PHEMA, the hydroxyl oxygen atoms in the acrylate
side chains readily exchange their H-bonding partners from polymers
to nearby water molecules, resulting in a pronounced plasticization effect.
In contrast, PVA contains a relatively large fraction of hydroxyl oxygen atoms that
do not participate in H-bonding in the dry state owing to its short
side-chain structure. 
Although the proportion of H-bonds formed with water molecules
increases gradually upon hydration, this fraction remains smaller than those
of PHEA and PHEMA up to a water content of 25 wt\%, indicating a weaker
plasticization effect.
Figure~S4 of the Supplementary Material shows
the water content dependence of the probability distribution of
H-bonding partners associated with water hydrogen atoms acting as
donors.
This provides a complementary perspective to that shown in
Fig.~\ref{fig:numhb-poh} and illustrates how the H-bonding partners
shift from polymer functional groups to water molecules as the water
content increases, particularly for polymers such as PEG, PMEA, PMEMA,
and PBA that do not contain hydroxyl groups.

As the dry-state $T_\mathrm{g}$ decreases, the dependence of $T_\mathrm{g}$ on water
content becomes weaker, indicating a limited plasticization effect. 
This trend is particularly evident for PEG, for which $T_\mathrm{g}$ exhibits
little change with increasing water content, in agreement with
experimental observations.~\cite{graham1989Interaction, hey1991Polyethylene,
huang2001Interaction, hatakeyama2007Cold}
Interestingly, the water content dependence of the glass transition
temperatures $T_\mathrm{g}$
for the seven polymers with markedly different dry-state $T_\mathrm{g}$ values,
shown in Fig.~\ref{fig:tg}, is consistent with experimental findings 
that, in polyol liquids, the number of H-bondable hydroxyl groups
correlates with $T_\mathrm{g}$ in aqueous
environments.~\cite{nakanishi2011Systematic, vandersman2013Predictions}
When the water content exceeds 50 wt\% and water becomes excessive in
the system, $T_\mathrm{g}$ converges to a nearly constant value of
approximately 280 K, as shown in Fig.~\ref{fig:tg}. 
This behavior can be interpreted as water becoming the dominant
component, thereby diminishing the distinct characteristics of each
polymer. 
It should be noted, however, that water contents of 50 wt\% or higher
represent virtual conditions in the MD simulations and do not correspond
to the water contents of real systems.

\begin{figure*}[t]
\centering
\includegraphics[width=0.8\textwidth]{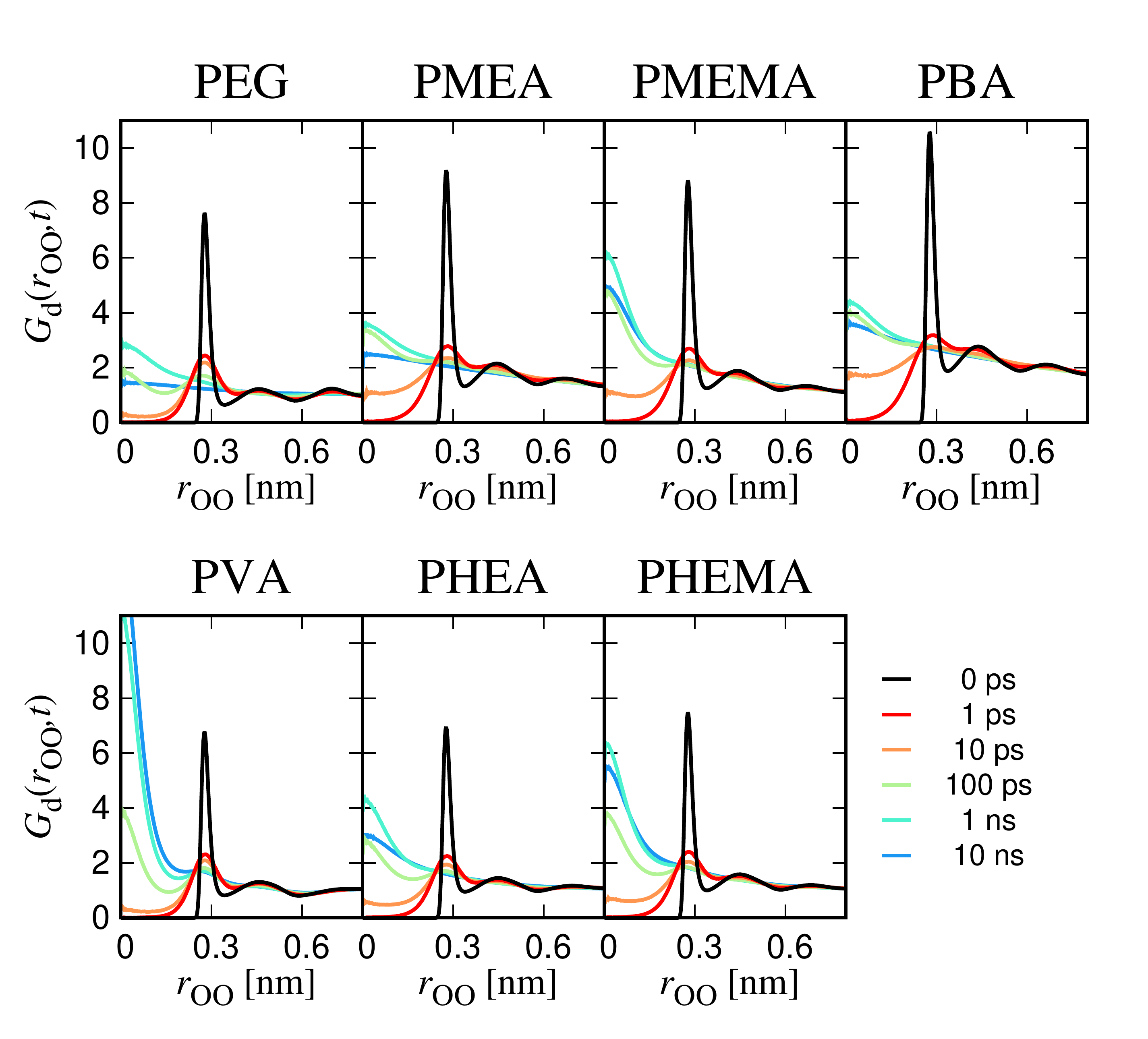}
\caption{Distinct part of the van Hove correlation function,
 $G_\mathrm{d}(r_\mathrm{OO}, t)$, 
for the oxygen-oxygen distance between water molecules at a water
 content of 25 wt\% and a temperature of 300 K.
The black curve ($t=0$ ps) corresponds to the radial distribution function between oxygen atoms.
}
\label{fig:vandww}
\end{figure*}

\subsection{Fluctuations in dihedral angles}
\label{section:fluctuation}

Section~\ref{section:tg} demonstrated that the glass transition
temperature $T_\mathrm{g}$ exhibits significant changes in its water content
dependence due to differences in side chains and functional groups.
Recently, Jin \textit{et al.} reported that fluctuations of dihedral
angles in polymer melts, including both main-chain and side-chain
degrees of freedom, correlate with $T_\mathrm{g}$.~\cite{jin2022Molecular, jin2023Computationally}
Building on these studies, we quantified dihedral angle fluctuations associated with the
conformations of polymer chains and examine their relationship with $T_\mathrm{g}$.
Here, we analyzed dihedral angles defined by the main chain carbon atoms
(C-C-C-C) as well as those involving oxygen substituents (O-C-C-O) at
the polymer terminal, as exemplified by PEG and PMEA.
Note that, in the case of PBA, the O-C-C-C dihedral angle, containing the
butyl group, is considered instead.

Figure~S5 of the Supplementary Material shows the
distributions of the dihedral angles for C-C-C-C and O-C-C-O at the
water content of 25 wt\%.
Similar overall trends were obtained at 10 wt\% and 50 wt\% and at other
temperatures; however, the peak height decreases with increasing temperature
and water content (data not shown).
For each polymer, the distributions exhibit peaks corresponding to the
gauche configuration (60$^\circ$ and 300$^\circ$) and the trans configuration (180$^\circ$).
Notably, the O-C-C-O dihedral angles of PEG, PMEA, and PHEMA
predominantly adopt the gauche conformation, indicating enhanced chain
flexibility.
Following the same algorithm described in
Ref.~\onlinecite{jin2022Molecular}, we employed a Gaussian mixture
model that represents the data as a superposition of Gaussian distributions.
Specifically, we approximated the distribution as a mixture of three
Gaussian components centered at the trans ($\approx 180^\circ$), gauche-
($\approx 60^\circ$), and
gauche+ ($\approx 300^\circ$) angles,
and extracted the corresponding center positions and standard deviations. 
The results are also shown in Fig.~S5 of the Supplementary Material.
The mixture proportion $w_i$ and standard deviation $\sigma_i$ of the $i$-th
Gaussian distribution were used to define the magnitude of the dihedral angle fluctuation as
$\sigma_\mathrm{E} = \sum_{i=1}^{3} w_i \sigma_i$.
The magnitude of $\sigma_\mathrm{E}$ quantifies the overall spread of the
dihedral angle distribution, and is therefore expected to correlate with
the extent of structural fluctuations within a given conformation as
well as the propensity for transitions between different conformations.

The temperature dependence of $\sigma_\mathrm{E}$ of C-C-C-C and O-C-C-O
angles for each polymer at 25 wt\% is plotted in Fig.~\ref{fig:sigma}.
As seen in the figure, 
$\sigma_\mathrm{E}$ decreases with decreasing
temperature. 
The magnitude of $\sigma_\mathrm{E}$ for 
C-C-C-C dihedral angle is relatively large for PMEA, PHEA, and PBA, and smaller
for PMEMA and PHEMA, which contain $\alpha$-methyl side group on the main
chain. 
PVA exhibits intermediate behavior. 
These results indicate that $\alpha$-methyl side groups along the
backbone reduce overall chain mobility. 
Furthermore, the temperature dependence of $\sigma_\mathrm{E}$ becomes more
gradual near and below the glass transition temperature, $T_\mathrm{g}$, 
consistent with 
previously reported findings.~\cite{jin2023Computationally}
To examine this in detail, 
the glass transition temperatures $T_\mathrm{g}^\mathrm{C-C-C-C}$ and
$T_\mathrm{g}^\mathrm{O-C-C-O}$ for the C-C-C-C and O-C-C-O dihedral angles,
respectively, are determined from the temperature
dependence of $\sigma_\mathrm{E}$ using the two linear fits, which is
shown in Fig.~\ref{fig:sigma}.
The relationship between $T_\mathrm{g}^\mathrm{C-C-C-C}$ and
$T_\mathrm{g}^\mathrm{O-C-C-O}$, and $T_\mathrm{g}$ in Fig.~\ref{fig:tg} is demonstrated
in Fig.~S6 of the Supplementary Material.
This figure shows that, except for PMEMA, which overestimates the $T_\mathrm{g}$
determined from density, a
correlation is observed for $T_\mathrm{g}^\mathrm{C-C-C-C}$
For $T_\mathrm{g}^\mathrm{O-C-C-O}$, a correlation is also present; however, all polymers
overall underestimates the $T_\mathrm{g}$ determined from density.

Figure~\ref{fig:sigma} also shows that the magnitude of $\sigma_\mathrm{E}$
for O-C-C-O (or O-C-C-C) dihedral angle decreases in the order PEG $>$ PMEA $>$ PHEA $>$
PHEMA $>$ PMEMA $>$ PBA.
In addition, PEG and PMEA exhibit a change in slope near $T_\mathrm{g}$,
analogous to that observed for the C-C-C-C dihedral angle.
PEG and PMEA exhibit greater mobility compared to other polymers,
suggesting that the rotational freedom of their ether groups results in
less steric constraint.
In contrast, in the other polymers, dihedral angle fluctuations are
reduced by steric constraints imposed by $\alpha$-methyl side chains and
by H-bonding involving hydroxyl oxygen atoms.
Note that 
the smallest value of $\sigma_\mathrm{E}$ observed for PBA originates
from fluctuations in the O-C-C-C dihedral angle. 
Compared with the O-C-C-O linkage, this behavior can be attributed to
the larger steric size and lower rotational freedom of the butyl group,
which restricts conformational flexibility.

\begin{figure*}[t]
\centering
\includegraphics[width=0.8\linewidth]{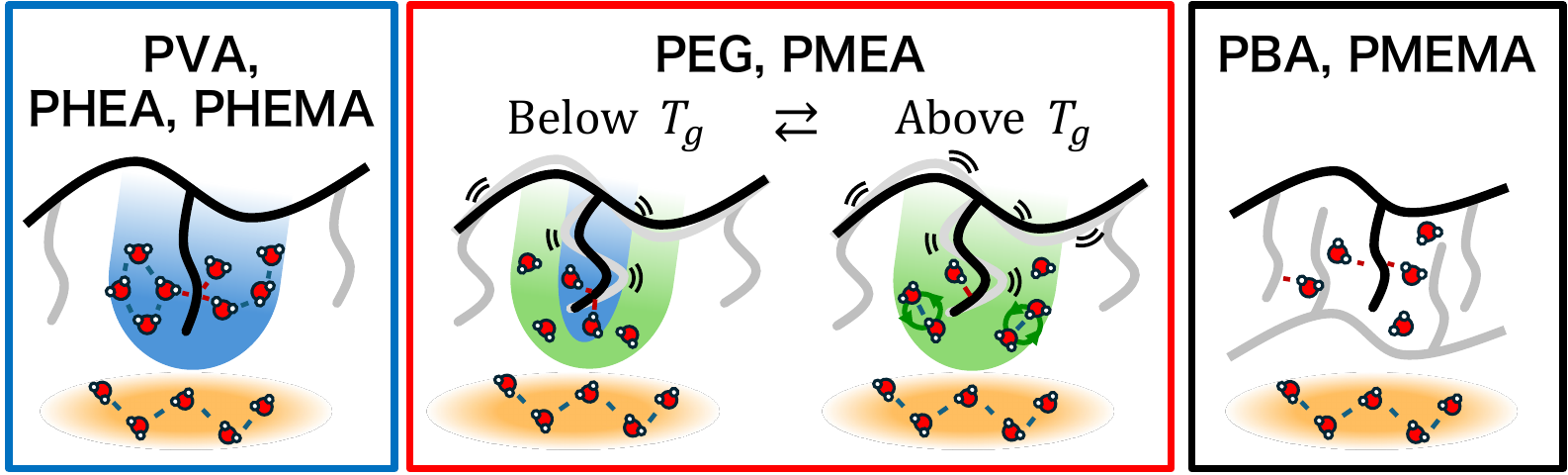}
\caption{Schematic illustration of the states of interfacial water in
 the vicinity of polymers.
Hydrated type (PVA, PHEA, PHEMA):
Low polymer chain mobility and strong H-bonding lead to the
 formation of a localized hydration shell, in which water molecules are
 tightly confined.
Flexibly hydrated type (PEG, PMEA):
Below $T_{\mathrm{g}}$, restricted polymer mobility results in strongly
 bound water H-bonded with ether or methoxy oxygen atoms,
 which exhibits dynamically distinct behavior from the surrounding, more
 mobile (loosely bound) water. 
Above $T_{\mathrm{g}}$, enhanced polymer mobility gives rise to a
 predominance of loosely bound water in the vicinity of the polymers.
Hydrophobic type (PBA, PMEMA):
Hydrophobic interactions induce aggregations of the
 polymers. 
The system is partitioned into polymer-bound water and bulk-like water.}
\label{fig:hydration}
\end{figure*}

\subsection{H-Bond lifetime}
\label{section:tauHB}

This subsection examines the dynamic behavior of water molecules in the vicinity
of the polymers.
For this purpose, we analyzed the H-bond time correlation function,
$P_\mathrm{HB}(t)$, using 10 ns trajectories in the $NVT$ ensemble 
over the temperature range 260 to 420 K at 20 K intervals.
Specifically, $P_\mathrm{HB}(t)$ represents the probability that an H-bond
remains intact at time $t$, given that a specific donor-acceptor pair is
H-bonded at $t=0$.~\cite{rapaport1983Hydrogen, luzar1996Effect,
luzar1996Hydrogenbond, luzar2000Resolving}
It is given as 
\begin{align}
P_\mathrm{HB}(t) = \frac{\ave{h_{i,j}(t)h_{i,j}(0)}}{\ave{h_{i,j}(0)}}, 
\label{eq:phb}
\end{align}
where
$h_{i,j}(t)$ is the H-bond indicator function, which takes a value of
unity 
if water molecule $i$ (acting as the donor) and the acceptor
oxygen $j$ belonging to a polymer functional group are H-bonded at
time $t$, and zero otherwise.
$\langle \cdots \rangle$ represents the ensemble average over all
possible pairs of H-bonds at the initial time 0.
The calculated $P_\mathrm{HB}(t)$ is shown in Fig.~S7 of
the Supplementary Material.
Note that $P_\mathrm{HB}(t)$ at 300 K for PMEA and PHEMA was reported in
Ref.~\onlinecite{shikata2023Revealing}.
$P_\mathrm{HB}(t)$ was fitted by the Kohlrausch--Williams--Watts  (KWW) function, 
$P_\mathrm{HB}(t) \approx \exp[-(t/\tau\kww) ^{\beta\kww}]$.
The integral of $P_\mathrm{HB}(t)$ characterizes the H-bond lifetime
$\tau_\mathrm{HB}$, 
\begin{align}
\tau_\mathrm{HB} = \int_0^\infty P_\mathrm{HB}(t) \mathrm{d} t, 
\end{align}
yielding an estimation of $\tau_\mathrm{HB}
=(\tau\kww/\beta\kww)\Gamma(1/\beta\kww)$ using the Gamma function $\Gamma(\cdot)$.

Figure~\ref{fig:tauhb} presents the Arrhenius plot of
$\tau_{\mathrm{HB}}$ as a function of inverse temperature for each
acceptor oxygen associated with the polymer functional groups at a 
water content of 25 wt\%.
As H-bond acceptors in the polymer functional groups, ether oxygen,
methoxy oxygen, hydroxyl oxygen, and carbonyl oxygen were considered. 
Here, ether-like oxygen atoms adjacent to carbonyl groups were excluded from
the analysis because their contribution to H-bond is negligible due to the
influence of the carbonyl group.~\cite{shikata2023Revealing}
We fitted the data to the Arrhenius equation, 
\begin{equation}
\tau_\mathrm{HB}  = \tau_0 \exp\left(\frac{E_\mathrm{A}}{RT}\right),
\end{equation}
where $\tau_0$ and $E_\mathrm{A}$ denote the characteristic time constant and the
activation energy associated with H-bond breakages, respectively, and
$R$ is the gas constant.
The straight lines in Fig.~\ref{fig:tauhb} represent the fitting
results in the high-temperature range (340-420 K).
Additional fittings are performed for ether oxygen of PEG, methoxy
oxygen of PMEA, hydroxy oxygen of PHEA, and hydroxy oxygen of PHEMA in
the low-temperature range (260-320 K).
The fitting parameters, $\tau_0$ and $E_\mathrm{A}$, are listed in
Table~S3 for the high-temperature range and
Table~S4 for the low-temperature range in the
Supplementary Material.
This suggests that in PVA and PEG, water molecules form H-bonds with
functional group oxygen atoms on the main chain; consequently,
structural constraints are significant, and H-bond breakage requires a
larger activation energy. 
In contrast, in acrylate polymers, H-bonding occurs with functional
groups on side chains that possess greater rotational freedom. 
Owing to this structural flexibility, the activation energy required for
H-bond breakage becomes lower.

At temperatures below the glass transition temperature $T_\mathrm{g}$,
a pronounced increase in $\tau_{\mathrm{HB}}$ is observed for both the
ether oxygen in PEG and the methoxy oxygen in PMEA, deviating from
Arrhenius behavior observed in the higher temperature regime.
In contrast, for polymer functional groups other than the ether oxygen
in PEG and the methoxy oxygen in PMEA, $\tau_{\mathrm{HB}}$ exhibits an
approximately Arrhenius temperature dependence.
These results suggest that the strength of H-bonding between water
molecules and the ether oxygen in PEG and the methoxy oxygen in PMEA
varies with temperature. 
In other words, above $T_{\mathrm{g}}$ the H-bonds are relatively weak and
transient, whereas below $T_{\mathrm{g}}$ they become more stable and
longer-lived.
For example, when comparing the methoxy oxygen in PMEA with those in
PHEA, PMEMA, and PHEMA, $\tau_{\mathrm{HB}}$ for PMEA is smaller at
high temperatures; however, at $T =
240$ K, below $T_\mathrm{g}$,
the values of $\tau_{\mathrm{HB}}$ become comparable.
Note that $\tau_{\mathrm{HB}}$ for hydroxy oxygen of PHEA and
PHEMA also shows slight deviations from the Arrhenius behavior, but the
magnitude of these 
deviations is smaller than those observed in PEG and PMEA, as evident in
$E_\mathrm{A}$ (see Table~S4 of the Supplementary Material).
This difference is attributed to the higher mobility of the methoxy
group compared with the hydroxyl group, as shown in
Fig.~\ref{fig:sigma}, resulting in a relatively lower
$T_\mathrm{g}$. 
Consequently, $\tau_{\mathrm{HB}}$ exhibits a marked change across
$T_\mathrm{g}$. 
Above $T_\mathrm{g}$, the high mobility of the methoxy group promotes H-bond
breakage with water, whereas below $T_\mathrm{g}$, the reduced mobility allows
H-bonds to persist for longer times.

\subsection{water molecule exchange dynamics}
\label{section:Gd}

Differences among functional group oxygens that form H-bonds with the
polymer are expected to influence the mobility of the associated water
molecules, particularly the exchange of positions between neighboring
water molecules.
To clarify this aspect, we analyzed the water molecule exchange dynamics via van Hove
correlation function, specifically its distinct part, 
$G_\mathrm{d}(r_\mathrm{OO}, t)$, defined in terms of the oxygen-oxygen
distance $r_\mathrm{OO}$ between two distinct water molecules.~\cite{hansen2013Theory}
Formally, $G_\mathrm{d}(r_\mathrm{OO}, t)$ is defined by
\begin{equation}
G_\mathrm{d} (r_\mathrm{OO}, t) = \frac{1}{4\pi r_\mathrm{OO}^2 \rho N}\left\langle \sum_{i=1}^N
					      \sum_{j\ne i}^N \delta
					      \left(r_\mathrm{OO}-|\bm{r}_j(t)
					      - \bm{r}_i(0)|\right)\right\rangle,
\end{equation}
where $\bm{r}_j(t)$ represents the position vector of the oxygen atom of the
$j$-th water molecule at time $t$.
Here, $\rho$ and $N$ denote the average number density and the number of water
molecules, respectively.
At time $t=0$, $G_\mathrm{d}(r_\mathrm{OO}, 0)$ is identical with the
oxygen-oxygen radial distribution function, $g(r_\mathrm{OO})$.
$G_\mathrm{d}(r_\mathrm{OO}, t)$ 
captures the relative spatial correlation of atom $j$ at time $t$ and a 
reference atom $i$ at time $t=0$, thereby providing insight into 
how local neighbor environments rearrange over time.
As demonstrated in Fig.~S8 of the Supplementary Material, 
$G_{\mathrm{d}}(r_\mathrm{OO}, t)$ for pure liquid water
shows that the amplitude
of $g(r_\mathrm{OO})$ decays monotonically with time and eventually
converges to unity regardless of $r_\mathrm{OO}$. 
This behavior indicates that structural relaxation leads to a loss of
memory of the initial molecular configuration, at times exceeding the
characteristic relaxation time, owing to 
frequent position exchanges between different molecules.

Figure~\ref{fig:vandww} shows the results for $G_{\mathrm{d}}(r_\mathrm{OO},t)$ at
a water content of 25 wt\% for each polymer.
For all polymers, the peak in $G_{\mathrm{d}}(r_\mathrm{OO},t)$
corresponding to the first hydration shell decreases progressively with time.
However, an increase in $G_{\mathrm{d}}(r_\mathrm{OO},t)$ is observed in
the vicinity of $r_{\mathrm{OO}} = 0$, a feature that has also been
reported for hydrated PVA systems.~\cite{muller-plathe1998Microscopic}
As discussed in Ref.~\onlinecite{muller-plathe1998Microscopic}, 
this behavior reflects the hopping motion of water molecules from the first
hydration shell toward the origin, which differs from that in pure water.
This 
suggests 
the development of spatial correlations 
between different water molecules over time, 
arising from water-water exchange
processes 
associated with the localization of water molecules confined within the
polymer matrix due to polymer-water interactions.
As shown in Fig.~\ref{fig:vandww}, 
this feature is 
particularly pronounced for 
PVA, PHEA, PMEMA, PHEMA, and PBA compared with PEG and PMEA.
PVA, PHEA, and PHEMA are hydrophilic polymers containing hydroxyl groups
and exhibit relatively long H-bond lifetimes with water, indicating a
strong confinement effect on the associated water molecules. 
In contrast, PMEMA and PBA are more hydrophobic and possess lower
equilibrium water contents, with a tendency for polymer chain
aggregation, as evident in Fig.~S1 of the Supplementary Material. 
Consequently, water molecules spatially confined within such polymer
matrices are expected to become localized.
By comparison, PEG and PMEA possess flexible backbones derived from ether
and methoxy groups, respectively, as demonstrated in Fig.~\ref{fig:sigma}. 
This flexibility enhances the mobility of confined water molecules,
resulting in a reduced localization effect as characterized by
$G_{\mathrm{d}}(r_{\mathrm{OO}}, t)$.

\section{CONCLUSIONS}

In this study, we performed MD simulations to systematically investigate
seven hydrated polymers, encompassing ethylene-based and acrylate
polymers. 
Specifically, we elucidated the behavior of interfacial water in each
polymer from four perspectives: (i) the water-content dependence of the
glass transition temperature, (ii) polymer chain fluctuations
characterized by dihedral angle distributions, (iii) H-bond lifetimes
between water molecules and polymer functional groups, and (iv) the
exchange dynamics of water molecules confined within the polymer matrix.

Based on these analyses, the seven polymers can be broadly classified into three groups.
The first group consists of hydrophilic polymers containing hydroxyl
groups, namely PVA, PHEA, and PHEMA. 
These polymers exhibit relatively high glass transition temperatures
$T_{\mathrm{g}}$ in the dry state and a pronounced decrease in
$T_{\mathrm{g}}$ with increasing water content. 
Their polymer chain fluctuations are small, and the H-bond lifetimes
$\tau_\mathrm{HB}$ between hydroxyl oxygen atoms and water molecules
follow Arrhenius behavior.
The second group comprises PBA and PMEMA, which are characterized by low
equilibrium water contents and hydrophobic properties. 
Their dry-state $T_{\mathrm{g}}$ values are slightly lower than those of
the hydrophilic polymers, and the reduction in $T_{\mathrm{g}}$ upon
hydration is less pronounced. 
Similar to the first group, however, they exhibit limited chain
fluctuations, and $\tau_\mathrm{HB}$ also obeys Arrhenius behavior.
The third group consists of PEG and PMEA, which contain ether and
methoxy groups, respectively. 
These polymers exhibit relatively low glass transition temperatures
$T_{\mathrm{g}}$ in the dry state, and the influence of water content on
$T_{\mathrm{g}}$ is modest. 
Owing to the rotational freedom of the ether and methoxy oxygen atoms,
polymer chain fluctuations are comparatively large. 
In addition, the H-bond lifetimes $\tau_\mathrm{HB}$ display
super-Arrhenius behavior below $T_{\mathrm{g}}$.

Furthermore, in conjunction with the localization effect of water
molecules confined within the polymer matrix, as characterized by the
van Hove correlation function, $G_\mathrm{d}(r_\mathrm{OO}, t)$, the
interfacial water surrounding the polymers can be classified according
to the hydration mechanism into three categories: 
highly hydrated (PVA, PHEA, and PHEMA), hydrophobic (PMEMA and PBA), and
flexibly hydrated (PEG and PMEA) types. 
A schematic illustration of this conceptual framework is presented in
Fig.~\ref{fig:hydration}.
For the highly hydrated type, each polymer chain maintains a stable
structure and forms a localized hydration shell through H-bonding
mediated by hydroxyl oxygen atoms.
For the hydrophobic type, the tendency of polymer chains to aggregate
due to hydrophobic interactions leads to a localization of
H-bonded water molecules.
For the flexibly hydrated type (PEG and PMEA), the polymer chains
exhibit high mobility owing to the presence of ether or methoxy oxygen
atoms, suggesting that water molecules H-bonded to these sites are only
loosely bound. 
However, below $T_{\mathrm{g}}$, the activation energy associated with
H-bond dissociation increases, indicating the formation of stronger
H-bonds under glassy conditions.
Although a direct correspondence with DSC measurements is not
straightforward, 
a consistent interpretation can be proposed for PEG and PMEA as follows. 
Below $T_{\mathrm{g}}$, water molecules located in the vicinity of ether
or methoxy oxygen atoms are constrained within the glassy polymer
matrix. 
Upon heating above $T_{\mathrm{g}}$, the recovery of polymer chain
mobility enhances the diffusivity of these water molecules, enabling
their transition to a diffusion-dominated regime. 
If this process leads to low-temperature crystallization during the
heating scan, it provides a picture that is consistent with the concept
of intermediate water.

As a final remark, 
it is worth noting that the conceptual picture in
Fig.~\ref{fig:hydration} incorporates dynamic information, including
dihedral angle fluctuations, 
H-bond lifetimes, and water molecule exchange dynamics.
In this sense, Fig.~\ref{fig:hydration} is not directly consistent with
the instantaneous static structure characterized by the radial
distribution function 
$G_\mathrm{d}(r_\mathrm{OO}, t=0)$ (Fig.~\ref{fig:vandww}), and the
simulation snapshots (Fig.~S1 of the
Supplementary Material).
In other words, 
the dynamic behavior of interfacial water cannot be inferred solely from
instantaneous static structural information and require explicit
consideration of dynamic properties.

\section*{SUPPLEMENTARY MATERIAL}

The supplementary material provides the experimental properties of the seven
polymers (Table~S1), the system equilibration protocol
(Table~S2), fitting parameters
of the Arrhenius equation (Table~S3 and
Table~S4), 
MD simulation snapshots (Fig.~S1), 
the temperature dependence of mass density
(Fig.~S2), 
relationship between the corrected glass transition temperature and the
experimental values (Fig.~S3), 
proportions of H-bonding partners for water hydrogen atoms (Fig.~S4),
distributions of the dihedral angles
(Fig.~S5), 
relationship between the glass transition temperatures determined from
dihedral angle fluctuations and from density
(Fig.~S6), 
the H-bond correlation function
$P_\mathrm{HB}(t)$ (Fig.~S7), 
and distinct part of van Hove correlation function $G_\mathrm{d}(r_\mathrm{OO}, t)$ for
pure water (Fig.~S8).

\begin{acknowledgments}
The authors thank Hidekazu Kojima for helpful comments.
This work was supported by 
JSPS KAKENHI Grant-in-Aid 
Grant Nos.~\mbox{JP25KJ1764}, ~\mbox{JP25K00968}, \mbox{JP24H01719}, \mbox{JP22H04542},
 \mbox{JP22K03550}, 
 and \mbox{JP23H02622}.
We acknowledge support from
the Fugaku Supercomputing Project (Nos.~\mbox{JPMXP1020230325} and \mbox{JPMXP1020230327}) and 
the Data-Driven Material Research Project (No.~\mbox{JPMXP1122714694})
from the
Ministry of Education, Culture, Sports, Science, and Technology
and by
Maruho Collaborative Project for Theoretical Pharmaceutics.
The numerical calculations were performed at Research Center of
Computational Science, Okazaki Research Facilities, National Institutes
of Natural Sciences (Projects: 25-IMS-C052 and 26-IMS-C051).
\end{acknowledgments}

\section*{AUTHOR DECLARATIONS}

\section*{Conflict of Interest}
The authors have no conflicts to disclose.

\section*{Data availability statement}

The data that support 
the findings of this study are available from
the corresponding author upon request.

%

%

\end{document}


\noindent{\bf\Large Supplementary Material}
\vspace{5mm}
\begin{center}
\textbf{\large Classification of interfacial water governed by water-polymer
interactions in hydrated polymers:\\
 A molecular dynamics simulation study
of ethylene-based and acrylate polymers
}
\\

\vspace{5mm}
 
{Atsuki Hashimoto, Kokoro Shikata, Kang Kim, and Nobuyuki Matubayasi}
\\

\vspace{5mm}

\noindent
\textit{
Division of Chemical Engineering, Graduate School of Engineering
 Science, The University of Osaka, Toyonaka, Osaka 560-8531, Japan}

\end{center}

\begin{table}[H]
    \centering
    \caption{The water solubility, equilibrium water content, glass transition
temperature in the dry state $T_\mathrm{g}$, blood compatibility, and presence or
absence of intermediate water for polymers examined in this study.}

\vspace{2mm}

    \begin{tabular}{cccccccc}
\toprule
\midrule
         & PEG & PMEA & PMEMA & PVA & PHEA & PHEMA &  PBA \\ 
\midrule
        water solubility & + & - & - & + & - & - & - \\ 
        equilibrium water content [wt\%]         &      & 9.0~\cite{tanaka2004Effect} & 4.2~\cite{tanaka2004Effect} &  & 60~\cite{tanaka2004Effect} & 39~\cite{tanaka2004Effect} & 1.9~\cite{tanaka2004Effect} \\
        glass transition temperature in the dry state $T_\mathrm{g}$ [$^\circ$C]  &
	 -51~\cite{trotzig2007Structure} & -35~\cite{sato2015Relationship} & 26~\cite{sato2015Relationship} & 85~\cite{hodge1996Free} & 15.2~\cite{salmeronsanchez2003Influence} & 115~\cite{roorda1988Thermal} & -47~\cite{sato2015Relationship}  \\
        platelet adhesion inhibition      & +~\cite{santos-martinez2014Pegylation}   & +~\cite{tanaka2004Effect}   & - & - & - & - & - \\
        intermediate water             & +~\cite{hatakeyama2007Cold}   & +~\cite{tanaka2004Effect}   & - & - & - & - & - \\
\midrule
\bottomrule
    \end{tabular}
    \label{table:tg_data}
\end{table}

\begin{table}[H]
    \centering
    \caption{System equilibration protocol. At step 4, the heating
 process of 300 $\to$ 500 K and the subsequent cooling process 500 $\to$
 300 K each required 100 ps.
This temperature cycling was repeated 10 times.}

\vspace{2mm}

    \begin{tabular}{ccccc}
\toprule
\midrule
        Step & Run & Time [ns] & Temperature [K] & Pressure [bar]  \\ 
\midrule
        1 & $NVT$ & 1 & 1000 & - \\ 
        2 & $NPT$ & 1 & 1000 & 1000 \\ 
        3 & $NPT$ & 1.5 & 1000 $\to$ 300 & 1000 \\ 
        4 & $NPT$ & 2 &  300 $\rightleftarrows$ 500 & 1 \\ 
        5 & $NPT$ & 10 & 300 & 1000 \\ 
        6 & $NPT$ & 10 & 300 & 1 \\ 
        7 & $NPT$ & 5 & 600 or 500  & 1 \\ 
\midrule
\bottomrule
    \end{tabular}
    \label{table:SI_md_step}
\end{table}

\begin{table}[H]
    \centering
    \caption{Fitting parameters, $\tau_0$ and $E_\mathrm{A}$, of the
 Arrhenius equation for H-bond lifetime, $\tau_\mathrm{HB} = \tau_0
 \exp(E_\mathrm{A}/RT)$, for oxygen atoms in polymer functional groups
 in the high-temperature range (340-420 K).}

    \begin{tabular}{cccccccccccc}
\toprule
\midrule
&  PEG & \multicolumn{2}{c}{PMEA} & \multicolumn{2}{c}{PMEMA} & PBA &
     PVA & \multicolumn{2}{c}{PHEA} & \multicolumn{2}{c}{PHEMA}\\ 
&  Ether & Carbonyl & Methoxy & Carbonyl & Methoxy & Carbonyl & Hydroxy
     & Carbonyl & Hydroxy & Carbonyl & Hydroxy \\
\midrule
$\tau_0\times 10^5$ [ps] & $1.18$ & $31.5$ & $8.10$ & $25.9$
     & $13.9$ & $14.2$ & $0.685$
      & $41.6$ & $20.0$ &
     $27.9$ &  $18.8$\\ 
$E_\mathrm{A}$ [kJ/mol] & $45.0$ & $34.9$ &
		 $35.0$ & $35.9$ & $36.3$ & $38.0$ & $48.8$ & $35.2$ &
     $35.0$ & $36.2$ & $35.1$\\
\midrule
\bottomrule
    \end{tabular}
    \label{table:SI_Arrhenius_high}
\end{table}

\begin{table}[H]
    \centering
    \caption{Fitting parameters, $\tau_0$ and $E_\mathrm{A}$, of the
 Arrhenius equation for H-bond lifetime, $\tau_\mathrm{HB} = \tau_0
 \exp(E_\mathrm{A}/RT)$, for oxygen atoms in polymer functional groups
 in the low-temperature range (260-320 K).}

    \begin{tabular}{ccccc}
\toprule
\midrule
&  PEG & PMEA & PHEA & PHEMA\\ 
&  Ether & Methoxy & Hydroxy  & Hydroxy \\
\midrule
$\tau_0\times 10^5$ [ps] & $2.94\times 10^{-4}$  &  $2.13\times 10^{-2}$
	     & $5.18\times 10^{-1}$ &  1.59 \\
$E_\mathrm{A}$ [kJ/mol] & 67.1 & 51.0 & 45.0 & 41.9 \\
\midrule
\bottomrule
    \end{tabular}
    \label{table:SI_Arrhenius_low}
\end{table}

\begin{figure}[H]
\centering 
\includegraphics[width=0.8\textwidth]{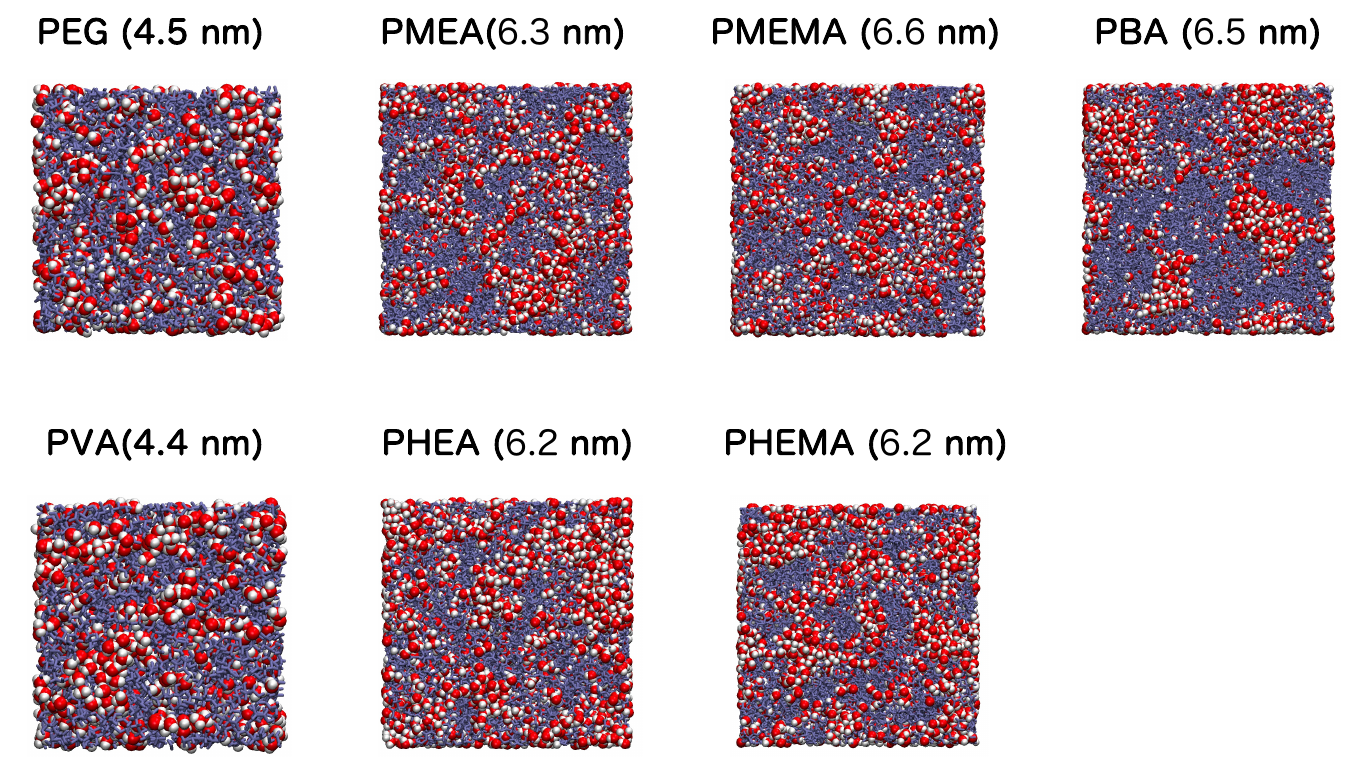}
\caption{MD simulation snapshots of the seven polymers at a water content
 of 25 wt\% and a temperature of 300 K.
Note that the size ratios of the individual images do not correspond
 to those of the MD simulation systems. 
The linear dimensions of each
 polymer system are indicated in parentheses.
}
\label{fig:SI_snapshot}
\end{figure}

\newpage
\begin{figure}[H]
\centering 
\includegraphics[width=0.8\textwidth]{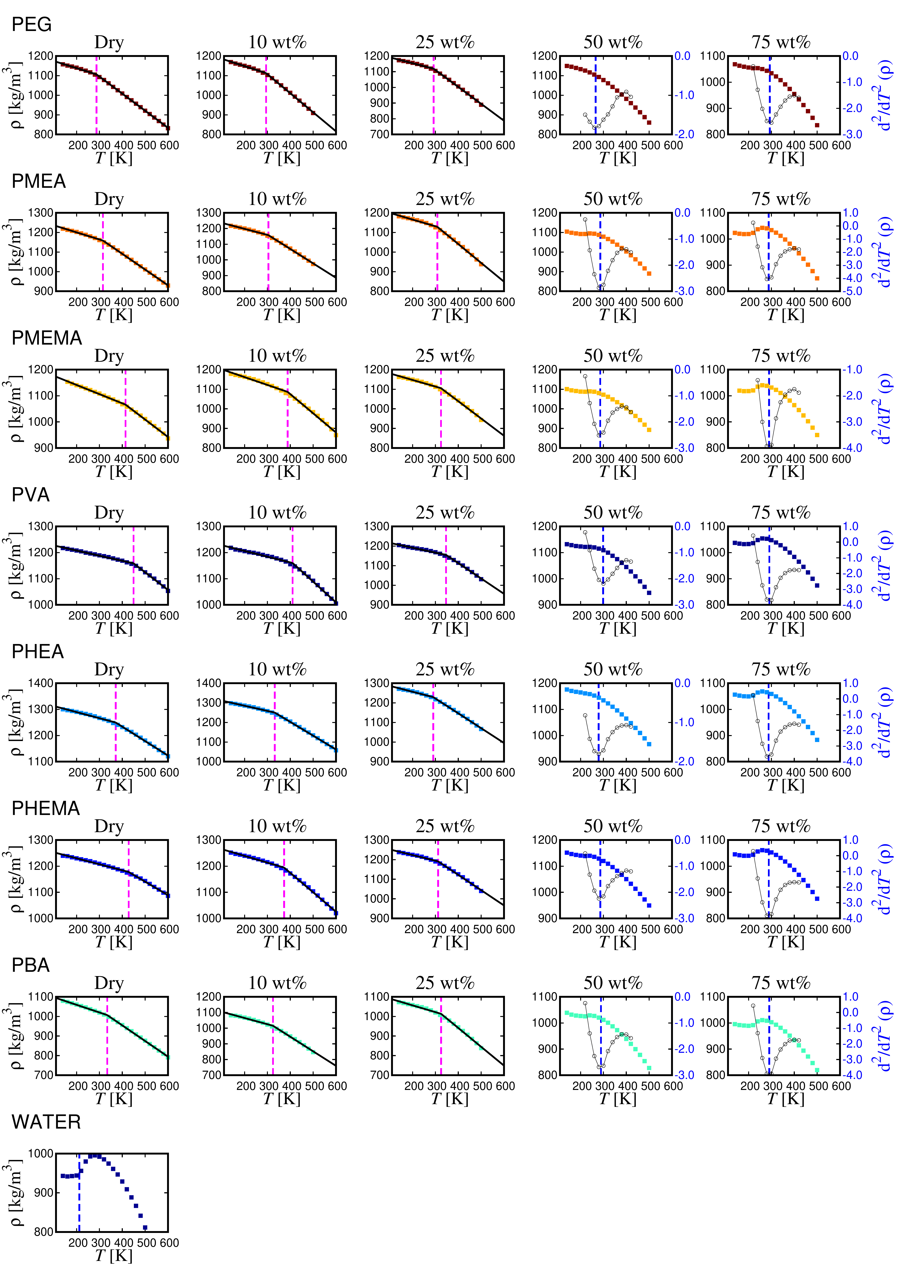}
\caption{Temperature dependence of mass density $\rho$ (left axis).
For water contents below 25 wt\%, $T_\mathrm{g}$ is determined from the
 intersection of two straight line fits (black lines), as indicated by
 the magenta dashed lines.
In contrast, for water contents above 50 wt\%, $T_\mathrm{g}$ is defined as the
 temperature at which the second derivative (right 
 axis), obtained from spline interpolation, reaches a minimum, as
 indicated by the blue dashed lines.
For pure water, the blue dashed line denotes the previously
 reported glass transition temperature of 212 K obtained from MD
 simulations.~\cite{kreck2014Characterization}}

\label{fig:SI_density}
\end{figure}

\newpage
\begin{figure}[H]
\centering 
\includegraphics[width=0.6\textwidth]{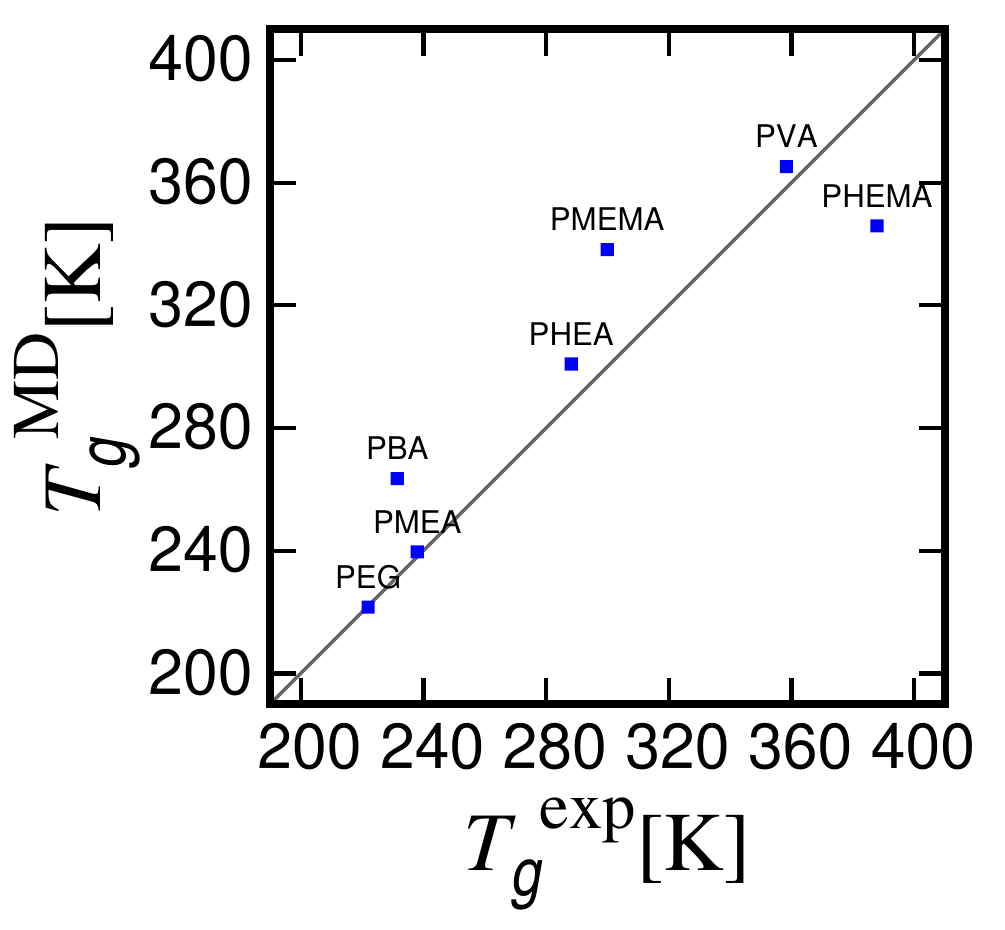}
\caption{Relationship between the corrected glass transition
 temperature $T_\mathrm{g}^\mathrm{MD}$ and the experimental values
 $T_\mathrm{g}^\mathrm{exp}$.
$T_\mathrm{g}^\mathrm{MD}$ is obtained using the Williams--Landel--Ferry equation.~\cite{soldera2006Glassa}
}

\label{fig:SI_Tg}
\end{figure}

\begin{figure}[H]
\centering 
\includegraphics[width=0.8\textwidth]{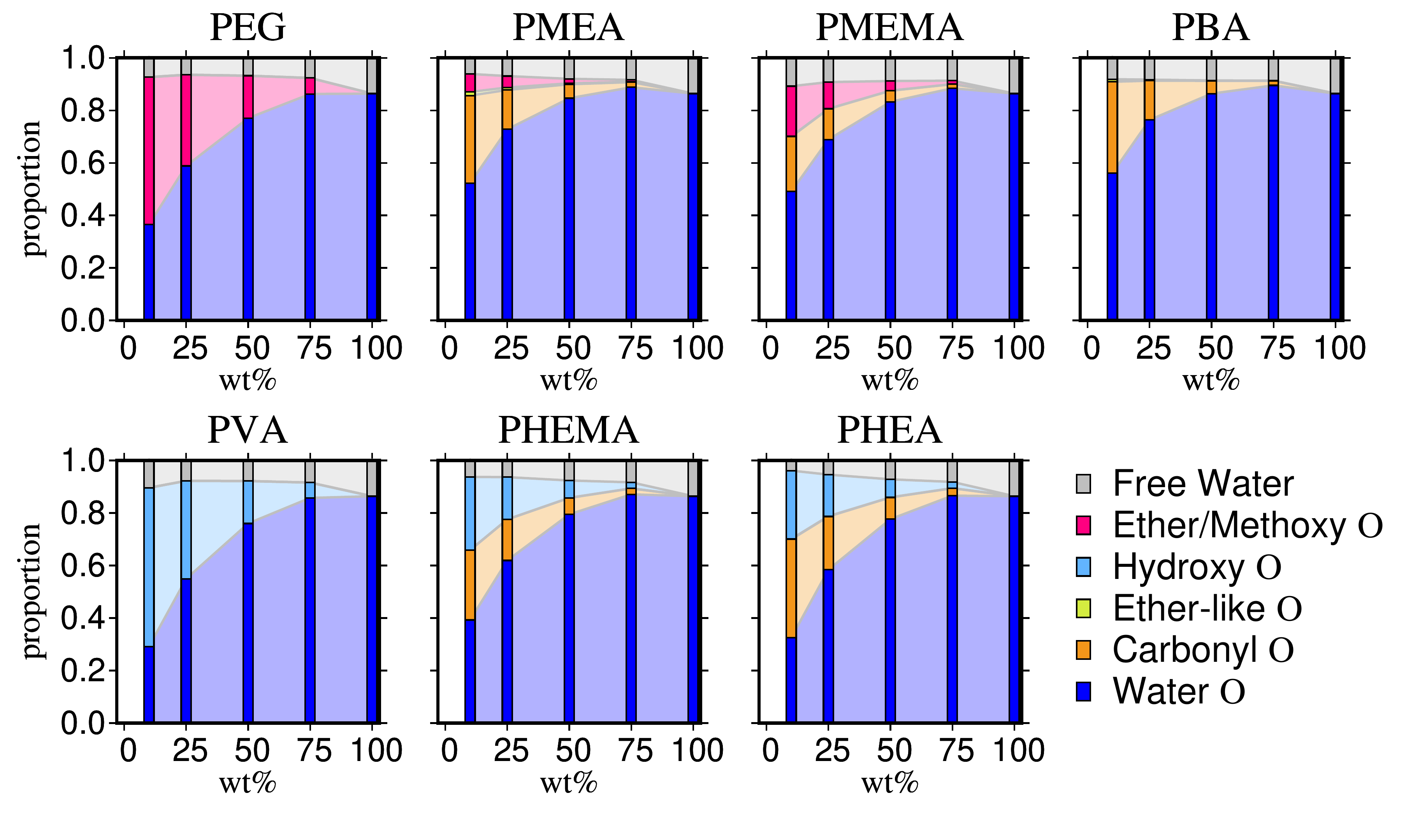}
\caption{Proportions of H-bonding partners for water hydrogen atoms at
 300 K.
Dark blue denotes H-bonds to water oxygen atoms; 
orange denotes H-bonds to carbonyl oxygen atoms; green
 denotes H-bonds to ether-like oxygen atoms; light blue denotes H-bonds
 to hydroxy oxygen atoms; magenda denotes H-bonds to
ether
 or methoxy oxygen atoms ; gray denotes
 water hydrogen that do not participate in H-bonding.
}

\label{fig:SI_Hbond}
\end{figure}

\newpage
\begin{figure}[H]
\centering 
\includegraphics[width=\textwidth]{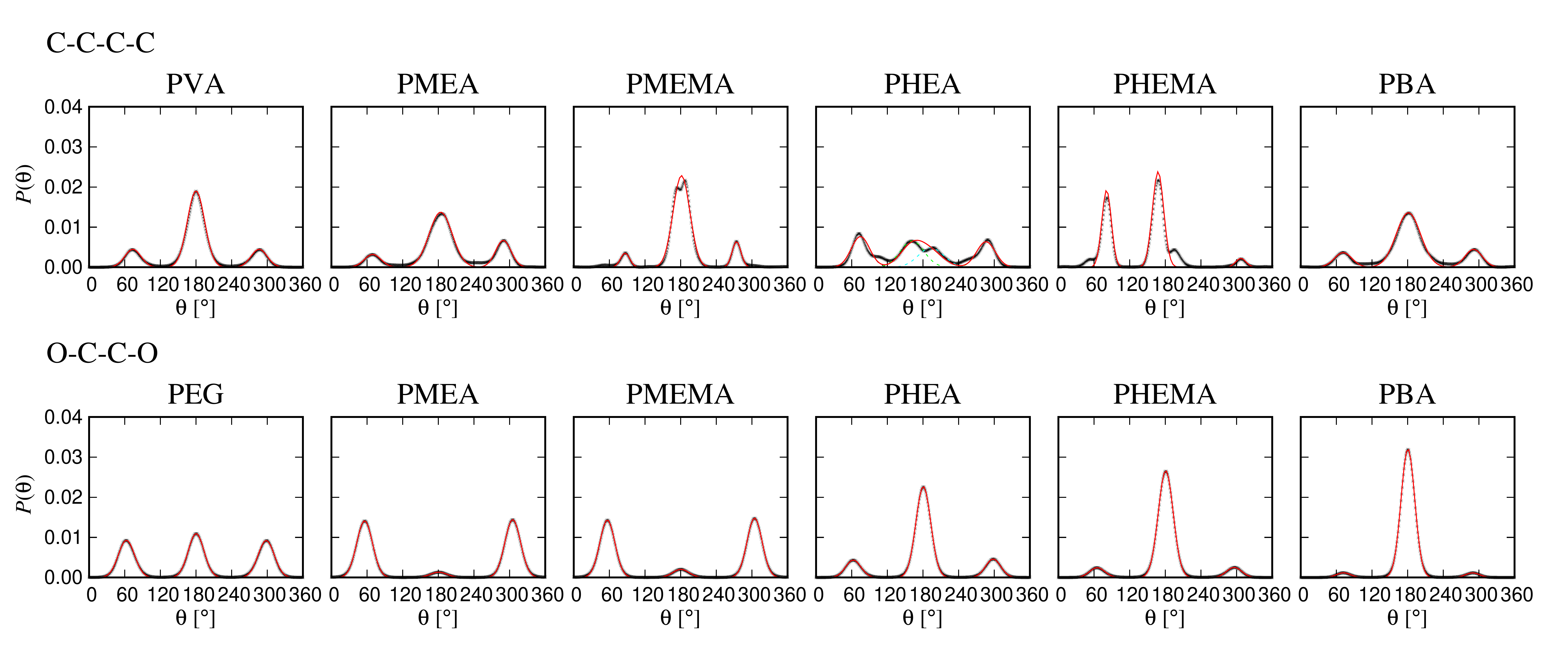}
\caption{Distribution functions of polymer dihedral angles: (top) dihedral
 angles of the main chain carbon backbone (C-C-C-C) and (bottom) dihedral
 angles of carbon skeleton substituted with oxygen atoms (O-C-C-O).
For PBA, the corresponding dihedral angle of the butyl side chain. 
Temperature: 300 K; water content: 25 wt\%. 
The black lines represent simulation data, and the red lines denote Gaussian mixture model fits.}

\label{fig:SI_Pd_dihhedral}
\end{figure}

\begin{figure}[H]
\centering 
\includegraphics[width=0.8\textwidth]{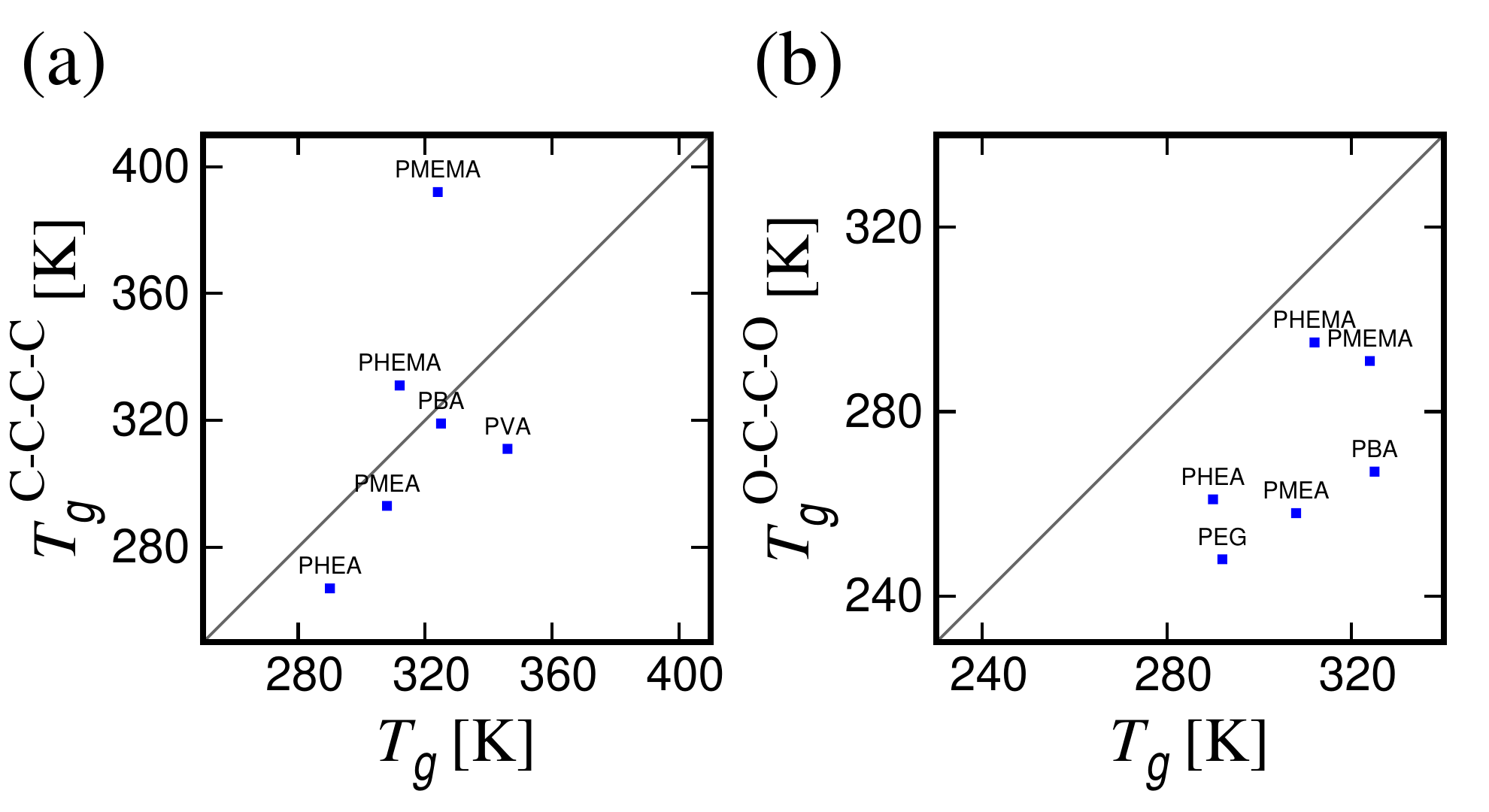}
\caption{Relationship between the glass transition temperatures
 $T_\mathrm{g}^\mathrm{C-C-C-C}$ (a) and $T_\mathrm{g}^\mathrm{O-C-C-O}$ (b) for the C-C-C-C and
 O-C-C-O dihedral angles, respectively, and $T_\mathrm{g}$ in Fig.~2
 of the main text.}

\label{fig:SI_Tg_dihedral}
\end{figure}

\newpage
\begin{figure}[H]
\centering 
\includegraphics[width=0.5\linewidth]{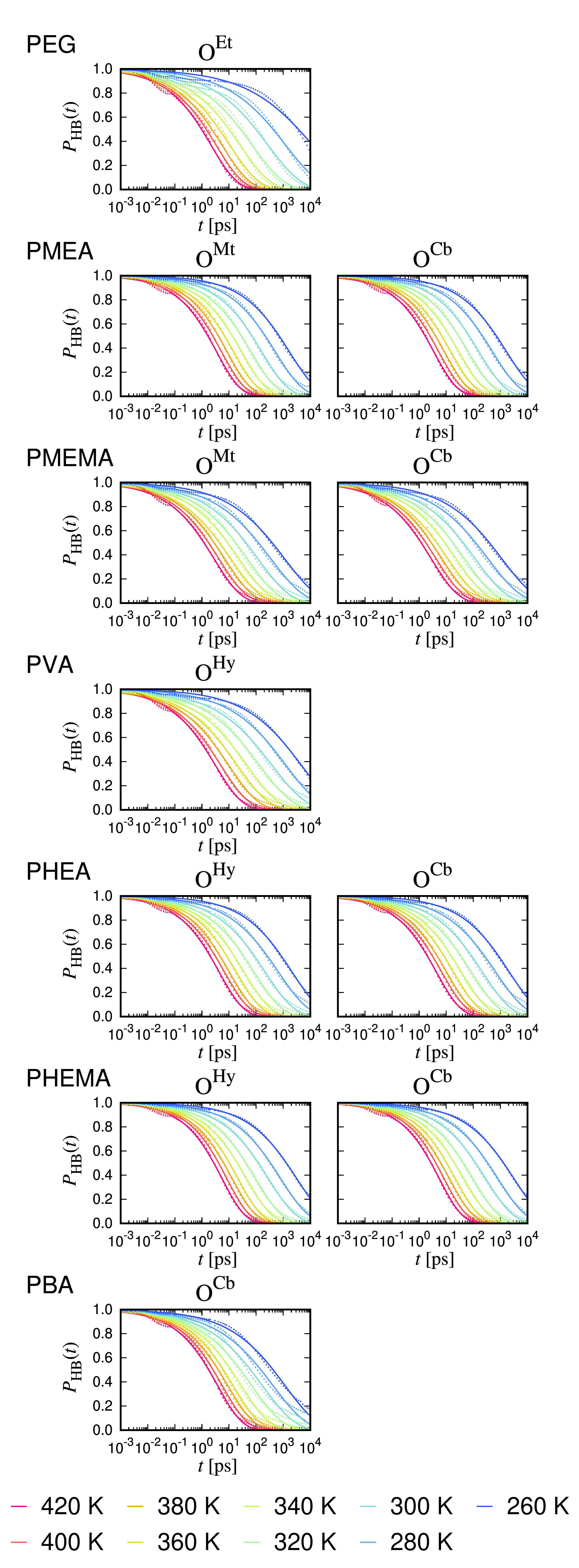}
\caption{H-bond time correlation function $P_\mathrm{HB}(t)$ obtained
 from 260 to 420 K at 20 K intervals.
The acceptor oxygen atoms involved in H-bonding with water molecules are
 denoted as ether oxygen (O$^\mathrm{Et}$), methoxy oxygen
 (O$^\mathrm{Mt}$), hydroxyl oxygen (O$^\mathrm{Hy}$), and carbonyl
 oxygen (O$^\mathrm{Cb}$).
Solid lines represent fits to the KWW function, 
$P_\mathrm{HB}(t) \approx \exp[-(t/\tau_\mathrm{KWW}) ^{\beta_\mathrm{KWW}}]$.}

\label{fig:SI_phb}
\end{figure}

\newpage
\begin{figure}[H]
\centering 
\includegraphics[width=0.5\linewidth]{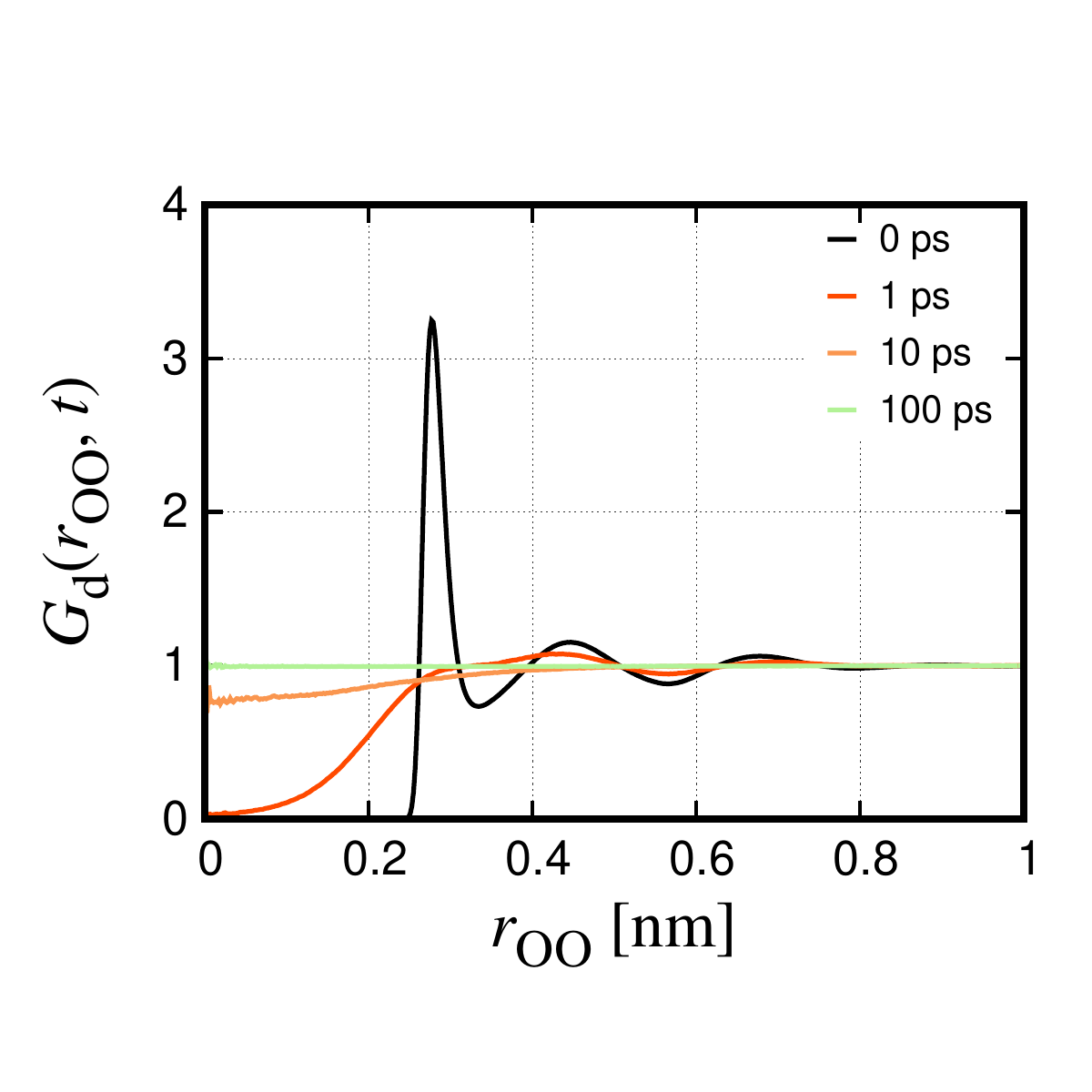}
\vspace*{-1cm}
\caption{Distinct part of van Hove correlation function
 $G_\mathrm{d}(r_\mathrm{OO}, t)$ for pure water at 300 K and 1 g/cm$^3$.
}

\label{fig:SI_Gd_water}
\end{figure}

%